\documentstyle[aaspp4]{article}

\input epsf
\newcommand{\sol}{$_{\scriptscriptstyle \odot}$}
\newcommand\Ang{${\rm\AA}$}
\newcommand\cc{\hbox{cm}^{-3}}
\newcommand{\eqn}[1]{(\ref{#1})}
\newcommand{\fig}[1]{\ref{#1}}
\newcommand{\msun}{\hbox{M}_\odot}
\newcommand{\Msun}{$\msun$}
\newcommand{\etal}{et~al.\ }
\newcommand{\gtaprx}{\gtrsim}
\newcommand{\ltaprx}{\lesssim}

\begin{document}

\title{A COMPARATIVE MODELING OF SUPERNOVA 1993J}

\author{S.I.Blinnikov\altaffilmark{1,2,3}
\affil{Institute for Theoretical and Experimental Physics,
 117259, Moscow, Russia; \\ email: blinn@sai.msu.su}}
\author{R.Eastman\altaffilmark{1}\affil{Lawrence Livermore National
        Laboratory, Livermore CA 94550;
        \\ email: eastman@tapestry.llnl.gov}}
\author{O.S.Bartunov\altaffilmark{1}, 
V.A.Popolitov\affil{Sternberg Astronomical Institute, 119899, Moscow,
          Russia;
        \\ email: oleg@sai.msu.su, vlad@sai.msu.su}}
\and \author{S.E.Woosley\altaffilmark{3,4}\affil{UCO/Lick Observatory,
 University of California, Santa Cruz, CA 95064;
\\ email: woosley@ucolick.org}
 }

\altaffiltext{1}{ UCO/Lick Observatory,
 University of California, Santa Cruz}
\altaffiltext{2}{Sternberg Astronomical Institute, Moscow, Russia}
\altaffiltext{3}{Max Planck Institut f\"ur Astrophysik, Garching,
 Germany, 85740}
\altaffiltext{4}{Lawrence Livermore National Laboratory, Livermore}

\lefthead{BLINNIKOV ET AL.}
\righthead{SN 1993J}

\date{\today}

\begin{abstract}

The light curve of Supernova (SN) 1993J is calculated using two
approaches to radiation transport as exemplified by the two computer
codes, STELLA and EDDINGTON. Particular attention is paid to shock
breakout and the photometry in the $U$, $B$, and $V$ bands during the first
120 days. The hydrodynamical model, the explosion of a 13 \Msun \ star
which had lost most of its hydrogenic envelope to a companion, is the
same in each calculation. The comparison elucidates differences
between the approaches and also serves to validate the results of
both.  STELLA includes implicit hydrodynamics and is able to model
supernova evolution at early times, before the expansion is
homologous.  STELLA also employs multi-group photonics and is able to
follow the radiation as it decouples from the matter.  EDDINGTON uses
a different algorithm for integrating the transport equation, assumes
homologous expansion, and uses a finer frequency resolution. Good
agreement is achieved between the two codes only when compatible
physical assumptions are made about the opacity.  In particular the
line opacity near the principal (second) peak of the light
curve must be treated primarily as absorptive even though the electron
density is too small for collisional deexcitation to be a dominant
photon destruction mechanism.  Justification is given for this
assumption and involves the degradation of photon energy by ``line
splitting'', i.e. fluorescence. The fact that absorption versus scattering matters to the
light curve is indicative of the fact that departures from equilibrium
radiative diffusion are important.
 A new result for SN 1993J is a prediction of the
continuum spectrum near the shock breakout (calculated by STELLA)
which is superior to the results of other standard single energy group
hydrocodes such as VISPHOT or TITAN. Based on the results of our
independent codes, we discuss the uncertainties involved in the
current time dependent models of supernova light curves.

\end{abstract}
\keywords{stars: supernovae -- radiative transfer -- hydrodynamics --
 methods: numerical -- stars: individual (SN 1993J)}

\section{INTRODUCTION}

Light curves are the most readily available data for diagnosing such
properties of supernovae as the mass of $^{56}$Ni ejected in the
explosion, the envelope mass, the asymptotic kinetic energy, and
evidence for mixing. Often direct comparison between models and data
is possible only for bolometric light curves, which makes them a
favorite of theorists. Much more data exists in the photometry of the
supernova in various wave bands, but models accurate enough for
comparison even to broad band observations require careful attention
to the sources of opacity and an accurate calculation of the
decoupling of radiation from the matter. Even the bolometric light
curve cannot be calculated, except very approximately, without the
same considerations.

In practice however, supernova light curves are often calculated using
the same crude approximations as employed in stellar evolution
codes. Typically, a single temperature is assumed for the matter and
radiation, and conditions at the photosphere are computed using
flux-limited radiation diffusion. An example of such a code is KEPLER
(Weaver, Zimmerman, \& Woosley 1978). The strength of such codes is
their ability to do nuclear physics, carry the star for extended
periods of time in hydrostatic equilibrium, simulate convection, and,
to some extent,  simulate the explosion mechanism. On the other hand,
the complex atomic physics of a supernova light curve is reduced in
such codes to a simple prescription for gamma-ray deposition and a
choice of a single opacity, usually electron scattering (assuming
thermal equilibrium) plus a constant.

At the other extreme, much greater attention may be paid to details of
the non-LTE (NLTE) atomic physics, and to the radiation transport, but
at the expense of hydrodynamic detail. Examples are the calculations
which have been performed by Swartz (1989), H\"oflich (1990), Eastman \&
Pinto (1993), and Baron \etal (1996b). Accurate radiation transport is
the strength of these codes, but their hydrodynamic ability is limited
to the coasting phase of the explosion, and they are very
computationally intensive.

Between these extremes, a variety of intermediate approaches are
possible. One such approach is embodied in the code STELLA (Blinnikov
\& Bartunov 1993, Bartunov et al. 1994). This code is able to treat,
in an implicit fashion, both the thermal and kinetic coupling between
gas and radiation field and to do so in both optically thick and thin
regimes.  The trade off required to achieve this ability is that
STELLA is currently limited by memory and floating point costs to a
relatively small number of photon energy groups.

STELLA is then able to treat shock propagation in a supernova (though
without nuclear reactions) and to calculate realistic light curves in
various colors. In particular, it is able to calculate the epoch of
shock breakout without any assumptions regarding the shape of the
spectrum (like blackbody radiation) as are required by single energy
group (or two-temperature) codes such as VISPHOT (Ensman \& Burrows
1992) or TITAN (Gehmeyr\& Mihalas 1994).  Since STELLA has not yet
been documented in the same detail as these other codes, we begin by
describing it (\S 2).

EDDINGTON (Eastman \& Pinto 1993), on the other hand, is a code
devoted more specifically to the radiation transport. The
hydrodynamics is very simple: free expansion is assumed.  In its most
intensive mode, EDDINGTON is capable of a full non-LTE simulation of
the spectrum (e.g. Eastman, Schmidt \& Kirshner 1996), but it may also
be run with a Saha-Boltzmann equation of state, which substantially
reduces the computational expense of computing a light curve. This is
because the computer costs in the non-LTE calculation are dominated by
solution of the atomic level populations. Here, this simplified
version of EDDINGTON is employed exclusively. Except for this one
approximation, all the physics of the non-LTE code is present - a
realistic prescription for $\gamma$-ray transport, fine energy
resolution, an opacity calculated using the line list of Kurucz
(1992), and a model for the effect of expansion on the opacity (see
\S\ref{OPACITY}).

The problem we chose to study with these codes is the light curve of
SN 1993J, a bright, well studied supernova (e.g. Shigeyama et
al. 1994; Woosley et al. 1994; Young, Baron \& Branch 1995; Utrobin
1996).  The specific stellar models are taken from Woosley et
al. (1994). In that paper some comparison was made between the light
curve predicted by single temperature flux limited diffusion in
KEPLER, and by multi-group transport in EDDINGTON. Here we concentrate
on the comparison with STELLA. The use of SN 1993J as a probe for the
theory is interesting since it was a peculiar Type IIb supernova which
had a low mass of hydrogen and is not so easy to model as a
standard Type II event.  On the other hand SN 1993J is not so hard to
model as Type Ia or Ib supernovae, which have virtually no hydrogen
and already show many indications of NLTE behavior near maximum light
(Lucy 1991; Eastman \& Woosley 1997). NLTE features of SN1993J do show
up, but well after maximum light (Utrobin 1996).

\section{THE CODE STELLA}

STELLA is an implicitly differenced hydrodynamics code that
incorporates multi-group radiative transfer.  The time-dependent
equations are solved implicitly for the angular moments of intensity
averaged over fixed frequency bands. The number of frequency groups
available for current workstation computing power, typically 100,
is adequate to represent, with reasonable accuracy, the non-equilibrium
continuum radiation.  For very large optical depths in the continuum
(typically, $\tau$ greater than 300), STELLA uses an equilibrium
diffusion approximation, similar to the transport scheme designed by
Nadyozhin and Utrobin (e.g., Litvinova \& Nadyozhin 1985; Utrobin,
1978, 1996).  Thus STELLA could, if necessary, be used even for
calculations of static stellar evolution (e.g., Blinnikov \&
Dunina-Barkovskaya 1994).

Instead of the intensity, $I_\nu$, STELLA works with the invariant photon
distribution function, that is the photon occupation number, $f_\nu$, where
\begin{equation}
 I_\nu={2h\nu^3\over c^2}f_\nu(r,\mu) \;
\label{inten}
\end{equation}
For spherical symmetry, $f_\nu$ is a function of the radial
distance, $r$, and of the cosine, $\mu$, of the angle between the
radius-vector, $\vec{r}$, and the direction of light propagation.

In a similar manner we introduce the blackbody photon distribution function
\begin{eqnarray}
 B_\nu(T)={2h\nu^3\over c^2}b_\nu(T), \quad \nonumber\\
  b_\nu(T)={1\over \exp(h\nu/k_{\rm B}T)-1}\;.
\label{bbinten}
\end{eqnarray}
If the angular moments of the distribution function, $f_\nu$, are defined:
\begin{eqnarray}
 {\cal J_\nu}={1\over 2}\int_{-1}^1 d\mu\,f_\nu; \quad\nonumber \\
 {\cal H_\nu}={1\over 2}\int_{-1}^1 d\mu \,\mu f_\nu;
 \quad\nonumber \\
 {\cal K_\nu}={1\over 2}\int_{-1}^1 d\mu \,\mu^2 f_\nu \;.
\label{momenf}
\end{eqnarray}
then after multiplication by $(2h\nu^3/c^2)$ one obtains the usual angular
moments of intensity, $I_\nu$
\begin{eqnarray}
  J_\nu={1\over 2}\int_{-1}^1 d\mu\,I_\nu \quad \nonumber\\
  H_\nu={1\over 2}\int_{-1}^1 d\mu \,\mu I_\nu  \quad\nonumber\\
  K_\nu={1\over 2}\int_{-1}^1 d\mu \,\mu^2 I_\nu \;
\label{momeni}
\end{eqnarray}
We can then rewrite the comoving-frame equations for the angular moments
(Castor 1972; Imshennik \& Morozov 1981;
Mihalas \& Mihalas 1984) in the form used in 
STELLA (the partial time derivatives  are taken for a fixed
Lagrangean coordinate variable):

\begin{eqnarray}
{\partial {{\cal J_\nu}}\over \partial t} = 
 -{c\over r^2}\cdot {\partial \over
\partial r}(r^2{\cal H_\nu})+c(\bar\eta_\nu-\chi_{\rm a} {\cal
J_\nu})+\nonumber\\
 +{u\over r}(3{\cal K_\nu}-{\cal J_\nu})
   -  {1\over r^2}\cdot
{\partial\over \partial r}(r^2u)
 ({\cal J_\nu}+{\cal K_\nu})\nonumber\\
 -{1\over \nu^3}
\cdot {\partial\over \partial \nu}
\nu^4\biggl[{u\over r}
(3{\cal K_\nu}-{\cal J_\nu})
 -{1\over r^2} \cdot {\partial\over \partial r} (r^2u){\cal K_\nu}
     \biggr]
\label{comov}
\end{eqnarray}

\begin{eqnarray}
{\partial {{\cal H_\nu}}\over \partial t} = 
  -c{\partial {\cal K_\nu}\over \partial r}-
{c\over r}(3{\cal K_\nu}-{\cal J_\nu})-\nonumber\\
 -2\biggl({u\over r}+{\partial u\over \partial r}
\biggr){\cal H_\nu}-c(\chi_{\rm a}+\chi_{\rm s}) {\cal H_\nu}
   +\dot{\cal H}_{\nu_{\rm diff}} \;.
\label{impul}
\end{eqnarray}
Here $u$ is the material velocity, $\chi_{\rm a}=\chi_{\rm
a}(\rho,T,\nu )$ is the absorptive opacity, and $\chi_{\rm s}$ is the
monochromatic scattering opacity (both having dimension of inverse
length, i.e. each is just the inverse of the corresponding mean free path).
The emission coefficient $\bar\eta_\nu=\chi_{\rm a} b_\nu(T)$ has the
same dimension and thus differs from the standard $\eta_\nu$.  The
current version of STELLA assumes that the emission coefficient
$\bar\eta_\nu=\chi_{\rm a} b_\nu(T)$.  Note that in this case the
standard source function $S_\nu \ne B_\nu$, since (monochromatic)
scattering is included.  So if we use a terminology like Mair \etal
(1992), this is already a NLTE effect taken into account by
STELLA, but we still use LTE Boltzmann-Saha distributions when
calculating ionization and level populations for $\chi_{\rm a}$ and
$\bar\eta_\nu$.
See the discussion in Blinnikov \& Bartunov
(1993) on the relative importance of the terms omitted and retained in
equations (\ref{comov}) and (\ref{impul}).
The term $\dot{\cal H}_{\nu_{\rm diff}}$, which provides
for artificial diffusion in the flux equation, is discussed in Appendix A.

Equations (\ref{comov}) and (\ref{impul}) are then solved
simultaneously for all
frequency groups using the
equations of hydrodynamics in Lagrangean coordinates,
\begin{equation}
 {\partial r\over \partial t} =u
\label{veloc}
\end{equation}
\begin{equation}
{\partial u\over \partial t} =4\pi
r^2{\partial (p+q)\over \partial m}-
 {Gm\over r^2}+a_{\rm r}+a_{\rm mix} 
\label{accel}
\end{equation}
\begin{equation}
{\partial r\over \partial m} ={1\over 4\pi r^2\rho} \;.
\label{conti}
\end{equation}
Here $p=p(\rho,T)$ is the material pressure, $\rho$, the density, $m$,
the Lagrangean coordinate (the mass inside radius $r$), $G$,
gravitational constant and $a_{\rm r}$, the radiative
acceleration:
\begin{equation}
 a_{\rm r}={4\pi \over c}\int_0^\infty (\chi_{\rm a}+\chi_{\rm s})
 {H_\nu\over \rho} d\nu \;.
\label{radac}
\end{equation}
The terms involving an artificial viscosity, $q$, and an additional 
acceleration, $a_{\rm mix}$, are discussed below.

We also require an equation for the material temperature, $T$, which is
obtained from (\ref{comov}), (\ref{conti}),
and the first law of thermodynamics:

\begin{eqnarray}
 \biggl({\partial e\over \partial T}\biggr)_\rho{\partial T\over \partial
 t} =\epsilon+4\pi \int_0^\infty
 \chi_{\rm a}{J_\nu-B_\nu \over \rho} d\nu -\nonumber\\
  - 4\pi {\partial{r^2u}\over \partial m}\biggl[T
 \biggl({\partial p\over \partial  T}\biggr)_\rho + q\biggr]\;,
\label{tempr}
\end{eqnarray}
with $e$, the specific internal energy of matter and $\epsilon$,
the specific
power of the local heating or cooling (for $\epsilon<0$).
In order to close the system (\ref{comov})--(\ref{tempr}) 
one must eliminate the
moment ${\cal K}$. As usual, we let
\begin{equation}
 {\cal K}=f_{\rm E} {\cal J} \;,
\label{closu}
\end{equation}
where $f_{\rm E}=f_{\rm E}(r,\nu)$ is the Eddington factor.
Once $f_{\rm E}$ is known
the system is closed and can be solved if appropriate
boundary conditions are given.

We assume that at the outer boundary ($m=M$, with $M$, the total mass)
the material pressure vanishes,
\begin{equation}
 p=0
\label{pboun}
\end{equation}
and there is no radiation coming from outside, which gives
\begin{equation}
 {\cal H}=h_{\rm E} {\cal J} \;.
\label{clobn}
\end{equation}
Here we introduce another Eddington factor $h_{\rm E}=h_{\rm E}(\nu)$.
The Eddington factors in equations  (\ref{closu}) and (\ref{clobn}) 
are found from a time-independent equation of radiative transfer (similar to
approaches of Herzig et al. 1990, Ensman \& Burrows 1992,
Gehmeyr \& Mihalas 1994, Zhang \&  Sutherland 1994, but now for
each energy group) using a procedure based upon the ideas of
Feautrier (1964). Instead of following the block-elimination
Feautrier scheme (see e.g. Mihalas 1978), STELLA uses the
elimination technique of Zlatev (see \O sterby \& Zlatev 1983)
which works for matrices having arbitrary patterns of sparseness.

In the equation of state,  $p=p(\rho,T)$, $e=e(\rho,T)$,
we take into account ionization and
recombination. The extinction $\chi=\chi(\rho,T,\nu)$ is
consistent with the equation of state
(see $\S$3). 

An important issue for the numerical treatment of shock break-out is
the need to introduce a special stabilizing term into
equation  (\ref{impul}). The reasons for this are discussed in the Appendix
A.

Two kinds of artificial accelerations are added to the physical
terms in equation  (\ref{accel}). The first, $q$, is the
gradient of the standard von Neumann viscous pressure used for
smearing shocks. The second, $a_{\rm mix}$, which we call the
``acceleration due to mixing'', is used to smear the thin dense layers
that appear in regions of thermal instability or at isothermal shocks
(Grasberg et al. 1971). Its form is given in Appendix B.

All space and frequency derivatives in 
equations  (\ref{comov}), (\ref{impul}), (\ref{accel}),
(\ref{conti}), and (\ref{tempr}) are replaced by finite differences. 
For the discretization STELLA uses up to 300 zones for the Lagrangean
coordinate and up to 100 frequency bins (for workstations
with RAM up to 128 MB), so a system
of ordinary differential equations (ODE) is produced for the evolution of
$r$, $u$, and $T$ in each Lagrangean zone and for ${\cal J}$ and ${\cal H}$
in each frequency group in each Lagrangean zone. This is
a version of the  method of lines (e.g. Oran \& Boris 1987)
which results here in tens thousands of ODE. This huge system of ODE 
is solved by an implicit high-order
predictor-corrector method based on the methods of Gear (1971) and
Brayton et al. (1972), see details in Blinnikov \& Panov (1996). 
The use of a scheme which is
implicit not only in the radiative transfer, but also in the
hydrodynamics allows one to overcome the Courant restriction
on the time-step and to apply the hydro-code not only
to dynamic problems, but also to situations where
hydrostatic equilibrium prevails.

The Eddington factors, $f_{\rm E}$, are computed separately for each
energy group in every mass zone.  To solve the large system of sparse
linear equations STELLA uses the technique of Zlatev (see \O sterby \&
Zlatev 1983) which is more flexible that the usual block elimination.
The latter approach is more difficult to deal with in situations when
the equations are changing during the run in various mass zones and
energy groups (e.g., for optically thick to thin cases), so here the
use of this technique is more crucial than in the usual solution of
the static radiative transfer equations.

STELLA calculates variable Eddington factors, but not every time
step. Following the method of quasi-diffusion (see Goldin 1964, and
Chetverushkin 1985, and references therein), Eddington factors are
computed only after a prescribed number of steps, $N_{\rm Edd}$
($N_{\rm Edd}=50$ in the current implementation). For 7000 -- 15000
steps required to calculate a typical light curve, the number of
computations of the Eddington factors compares well with $\sim 200$
large time steps in an EDDINGTON run.  One can say that the solution
of hydro and energy equations by STELLA for a number of small time
steps between consequent computations of Eddington factors is just
another algorithm for finding the temperature correction in EDDINGTON
(an evolutionary physical relaxation procedure).  For slow changes in
flux this is justified by the absence of any appreciable variations of
Eddington factors since the ratio of moments changes much less than
the moments themselves. For shock break out, we present
our arguments in the Appendix A.  The variable Eddington factors are
computed with full account of scattering and redshifts in
each energy group and employing an extrapolation of the Eddington
factors in time for the next large step.

STELLA computes the gamma-ray transfer in a one-group approximation. The
results for this non-local deposition agree quite well with
the independent algorithm used in EDDINGTON (Fig. \ref{gdeplg}).

The most important improvement in the current version of STELLA
(compared to that described by Blinnikov \& Bartunov, 1993) is the use
of tables of realistic opacities with expansion effect in lines taken
into account according to Eastman \& Pinto (1993). We now discuss this
in detail.

\section{OPACITY}
\label{OPACITY}

One of the important lessons learned in comparing the light curves
produced by STELLA and EDDINGTON was how sensitive the results were to
the line opacity, especially whether the opacity was treated as
scattering or absorptive. The effect of expansion on opacity has been
studied for a long time going back to work on the winds of Wolf Rayet
and early main sequence stars (e.g. Lucy 1971; Castor,
Abbott \& Klein 1975), as well as supernovae (Karp et al. 1977;
Wagoner, Perez, \& Vasu 1991; Eastman \& Pinto 1993; Blinnikov 1996,
1997, Baron \etal 1996a).
Even so, the brute force multi-frequency transport calculation remains
too expensive computationally because of the same difficulties which
plague line-blanketed stellar atmosphere calculations, namely the need
to resolve the frequency variation of each line in order to accurately
compute the effective photon transmittance.  Unfortunately,
approximate methods such as opacity distribution functions (e.g. Carbon 1979)
and opacity sampling (Peytremann 1974)
are not applicable to
hypersonic flows.  Some approximation must be adopted and in this
section we describe ours. A comparison and analysis of various
approximations to the expansion opacity has been given recently by
Pinto and Eastman (1997).

Our total opacity included contributions from photoionization,
bremsstrahlung, lines, and electron scattering. The bound-free
photoionization cross-sections were taken from Verner \& Yakovlev
(1995), who give analytic fits to their valence shell ground state and
inner shell photoionization cross-sections.
The line opacity used initially only in EDDINGTON, but ultimately in
both codes, was computed using atomic data from the line list of
Kurucz (1991). Approximately 110,000 lines were included. 
The line contribution to the total opacity was
computed using the approximation 
derived by Eastman \& Pinto (1993).  In this approximation the
opacity contribution from lines in a given frequency interval
$(\nu,\nu+\Delta\nu)$ is given by
\begin{eqnarray}
\chi_{\rm exp}={\nu\over\Delta\nu}{v\over
rc}\sum_j\int_0^1(1+Q\mu^2)\cdot\nonumber\\
\cdot \left\{1-\exp\left[-\tau_j(\mu)\right]\right\}\, d\mu
\label{expopac}
\end{eqnarray}
where the sum is over all lines in the interval $(\nu,\nu+\Delta\nu)$,
$Q\equiv d\ln v/d\ln r -1\approx 0$, and $\tau_j(\mu)=(h/4\pi) 
(n_l B_{ll^\prime}-n_{l^\prime}B_{l^\prime l})/ [(\partial v/\partial
r)(1+Q\mu^2)/c]$ is the Sobolev (1947) optical depth of line $j$ linking
levels $l$ and $l^\prime$, with level populations $n_l$ and Einstein
$B$ coefficients $B_{ll^\prime}$.

Eastman \& Pinto (1993) argued that the amount of time a
photon spends in resonance with a given line is much less than the
time spent out of resonance and in the continuum. For those photons which
scatter, the process of going into resonance with a line and
scattering around inside the line before finally being re-emitted on
the long wavelength side of the line and in a new direction can be
treated as a single scattering event. That is the meaning of
equation~\eqn{expopac}.  The effective opacity from lines in an interval
$(\nu,\nu+\Delta\nu)$ is the average number of line interactions
undergone while Doppler shifting through $\Delta\nu$, divided by the
distance traveled $\sim ct\Delta\nu/\nu$.

The density dependence in equation~\eqn{expopac} is unusual in that, for
optically thick lines ($\tau_j>>1$), $\chi_{\rm exp}$ is independent
of mass density, and proportional instead to the density of lines per unit
frequency interval.  For optically thin lines ($\tau_j<<1$) $\chi_{\rm
exp}$ behaves like other contributions to the opacity and scales
linearly with the density.

Instead of the expansion opacity parameter $ s=\chi_c ct$ (Karp et al.
1977, see also Blinnikov 1996, 1997) which is more appropriate for a large
optical depth in the continuum, the Eastman-Pinto expansion opacity
(\ref{expopac}), which is also applicable for negligible extinction in
the continuum, $\chi_c$, depends only on the value of $v/r$.  During
the coasting stage of free expansion the parameter $v/r$ is simply $1/t$ where
$t$ is the time after the explosion to high accuracy (it is the same
$t$ that enters the definition of Karp's parameter, $s$). At other
stages of explosion $v/r$ may differ significantly from the time $t$,
so we will denote $v/r$ in (\ref{expopac}) as $1/t_s$. Clearly the expansion
effect is higher for smaller $t_s$.

The reason we used the Eastman-Pinto approximation is chiefly because
it is simple and straightforward to derive. It also has a number of 
attractive properties, not the least of which is that it is a linear 
sum of the contributions from each line in a wavelength interval and 
therefore it can be added directly to the continuum opacity,
and is identical to the effective opacity derived by Castor, Abbott \& Klein (1975) 
for the free streaming limit).

Finally, the chief goal of the present work is to compare the
results of the codes STELLA and EDDINGTON. This requires the use
of a consistent choice of opacity,
the coding of which is already available. By calibrating against EDDINGTON
we also help to normalize and validate the results of 
Woosley et al. (1994).

\begin{figure}
\plotone{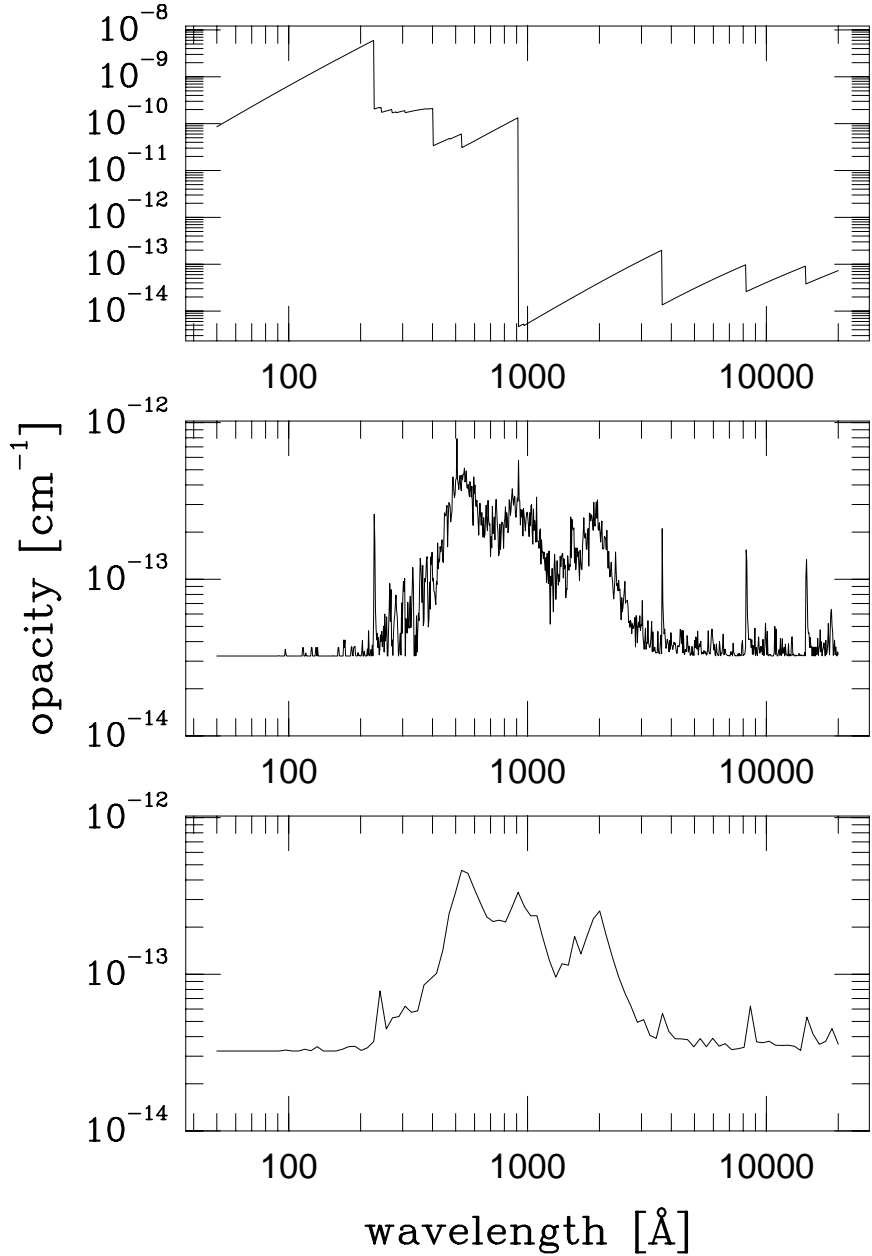}
\caption{Contributions to the total mass opacity from photoionization and
bremsstrahlung (top), and electron scattering and lines (middle),
for a solar composition, density
$\rho=10^{-13}\ \cc$, temperature $T=15,000$~K, time $t=15$~days, on a
frequency grid of 1000~points. In the bottom frame the sum of electron
scattering and line opacity is shown on a reduced frequency grid of
100~points.}
\label{opfig}
\end{figure}

To illustrate, Figure~\ref{opfig} displays the opacity contributions for
a solar composition at $\rho=10^{-13}\ \cc$, $T=15,000$~K, and time
$t_s=15$~days, on a frequency grid of $N_{\rm freq}=1000$ points. The
top panel shows the sum of bound-free and free-free opacities; the
lower panel gives the sum of the bound-bound and electron scattering
opacities. The strongest bound-free edges in the upper panel are from
hydrogen and helium. Most of the line opacity (lower panel) comes from
lines of Fe and other iron peak elements. The effect of the expression
~\eqn{expopac} is to represent the line opacity as a smooth
distribution. The bottom panel of Figure~\fig{opfig} shows the same
opacity distribution as in middle panel, but using one tenth as many
energy bins, and it can be seen by comparing the two that the
expansion opacity approximation (eq. [\ref{expopac}]) is insensitive to
the wavelength or frequency resolution in regions where the line
density is very large.  The EDDINGTON calculations used a frequency
grid of 500~energy bins, while STELLA employed a frequency grid of
100~points, but both grids accurately represented the ultraviolet line
opacity.
The one hundred frequency bins used by STELLA are adequate for the present work 
which is not an attempt to calculate accurate spectra, but only
broad band photometry. The advantage which STELLA has over other codes 
is its ability to solve implicitly coupled hydrodynamics and multi-group radiation transport.

For calculations with STELLA, opacity tables were produced for a range
of $T$, $\rho$, and velocity gradient, for a standard wavelength grid
spanning the range 10 to 50,000~\Ang, for $\sim 50$ compositions
characteristic of the range found in a given explosion model (the
composition of each 4th zone in a 200 zone representation of the
explosion model and of each zones in the 80 zone remap of the model
used by EDDINGTON).  For the opacity tables the Saha equation is
solved for the 6 most abundant ionization states of elements.  For
hydrodynamic runs STELLA uses another equation of state with only the 3
first ionization states computed accurately from Saha equation
following Karp (1980) and higher ionizations are treated in an
approximation of an averaged ion.  This equation of state is very
fast, smooth and gives all the necessary thermodynamic
derivatives. Tests showed no large differences for $N_e$ computed
using both methods.

\vskip 1 true cm

\section{STELLAR AND SUPERNOVA MODELS}

Since STELLA does hydrodynamics as well as radiation transport, we
compare its hydro capabilities here to another code, KEPLER (Weaver,
Zimmerman \& Woosley 1978), for the same presupernova star. KEPLER
does detailed hydrodynamics on a fine mass grid, but only crude
radiative transfer. The codes might be expected to agree so long as
the supernova remains very optically thick. Various models and runs
are summarized in Table 1.  
In Table \ref{runs},  $N_{\rm zon}$ \& $N_{\rm freq}$ are
the number of radial mass zones and frequency bins, respectively.  When
the entry for ``forced $\chi_{\rm abs}$'' is ``yes'', $\chi_{\rm abs}$
was artificially set equal to the total extinction (only in
this case does the source function is $S_\nu=B_\nu$).  When ``$\chi_{\rm
exp}$'' is marked with ``no'' it means that the opacity tables are
used with the parameter $t_s=100$ days, when the expansion effect on
the line opacity is very weak. If $t_{\rm start}=0$ then the full
explosion was computed by STELLA, otherwise the hydrodynamic structure is
the same as EDDINGTON, calculated by KEPLER for the given time after
core collapse.


\begin{table}
\caption{Runs\label{runs}}
\begin{center}
\begin{tabular}{lllllll}
\tableline
\tableline
run & $N_{\rm zon}$ & $N_{\rm freq}$ & forced $\chi_{\rm abs}$  
    & $\chi_{\rm exp}$ & $t_{\rm start}$, s  \\
\tableline
13C1  & 200 & 100 &  no    & no  & 0 		      \\
13C2  & 200 & 100 &  yes   & no  & 0 		      \\
13C2s  & 200 & 20  &  no    & no  & 0 		      \\
13C3  & 200 & 100  &  lines & no  & 0 		      \\
13C4  & 80  & 100 &  no    & no  & $2.05\times 10^5$  \\ 
13C5  & 80  & 100 &  lines & no  & $2.05\times 10^5$  \\ 
13C6  & 80  & 100 &  yes   & no  & $2.05\times 10^5$  \\ 
13C7  & 80  & 100 &  yes   & yes & $2.05\times 10^5$  \\ 
13C8  & 80  & 100 &  lines & yes & $2.05\times 10^5$  \\ 
\tableline
\end{tabular}
\end{center}
\end{table}

Blinnikov, Nadyozhin \& Bartunov (1991), Blinnikov \& Bartunov (1993)
and Bartunov et al (1994) all studied radiation transport in
supernovae using very simple structures for the presupernova star,
basically artificial constructions in hydrostatic equilibrium with the
necessary properties to reproduce an observed light curve.  The
current version of STELLA was modified to use realistic presupernova
models.  To run an explosion in reasonable time with the number of
energy groups employed by STELLA (20 -- 100), one can only use about
200 mass zones. This is less than used in KEPLER (300 -- 500). To map
the KEPLER model onto the STELLA grid, we used a modified version of
a code developed by Nadyozhin and Razinkova (1986) for constructing
initial hydrostatic models. In particular, it was necessary to have a
fine grid in the outermost layers (to enable non-equilibrium radiative
transfer there) and at the base of hydrogen envelope, especially for
such low hydrogen masses as in SN 1993J.

\begin{figure}
\plotone{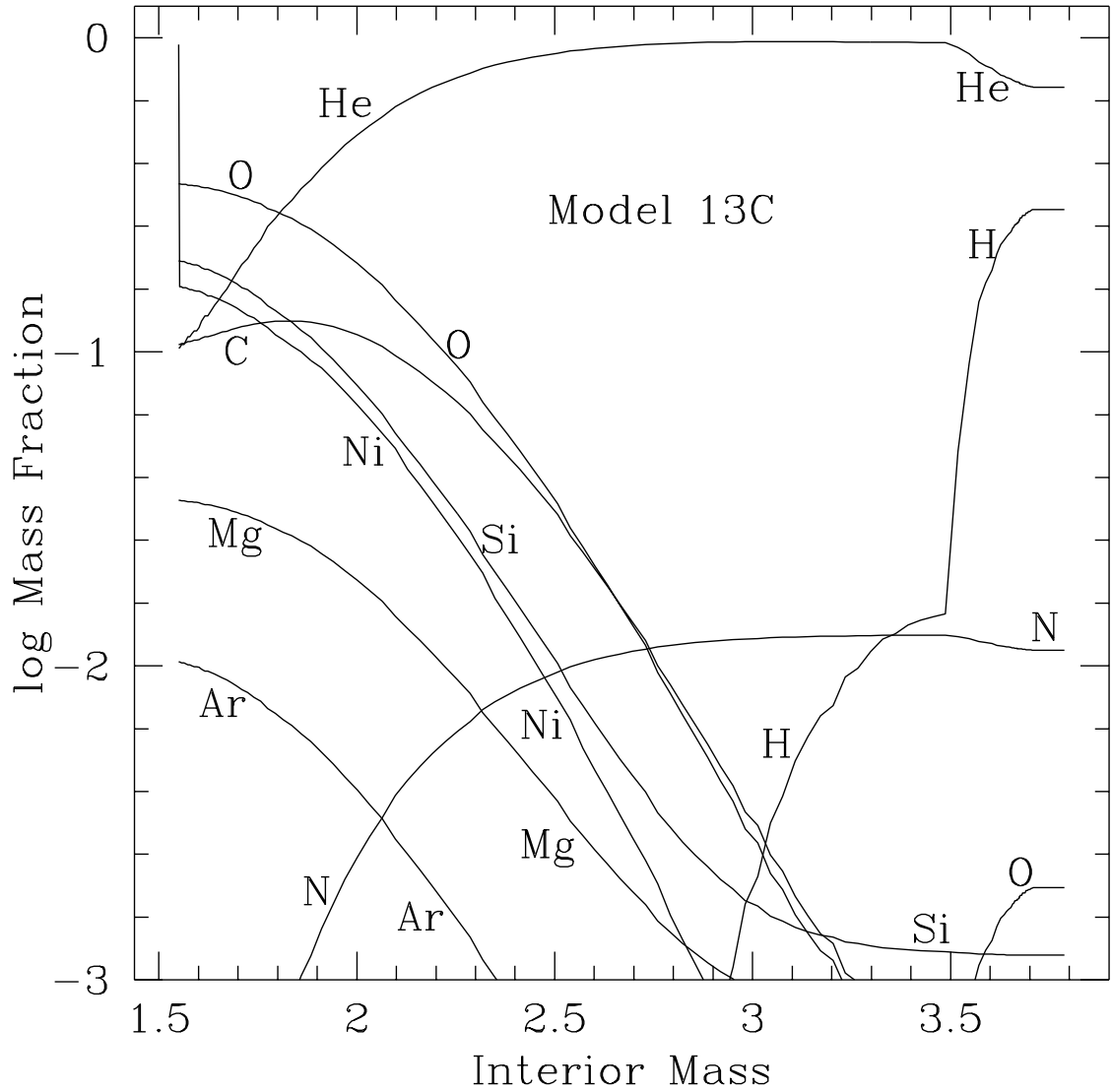}
\caption{ Final composition of Model 13C following explosive
     nucleosynthesis and a moderate amount of artificial mixing.}
\label{comp}
\end{figure}

The specific model employed here was Model~13C of Woosley et
al. (1994). This model was derived from a 13 M\sol \ main sequence
star that lost most of its hydrogen envelope to a nearby companion (9
M\sol\ initially 4.5 AU distant). The final presupernova star had a
helium and heavy element core of 3.71 M\sol, a low density hydrogen
envelope of 0.2 M\sol, and a radius of $4.33 \times 10^{13}$ cm. The
structure and composition of this model is similar to Model
13B shown in Figures 2 and 3 of Woosley et al. Explosion was simulated
in KEPLER by a piston at the outer edge of the iron core at
1.41 M\sol. This piston was first moved in briefly to simulate core
collapse then rapidly moved outwards to create a shock that ejected
most of the material external to the piston with high velocity. About
0.05 M\sol \ fell back onto the piston, presumably to become part of
the neutron star. 
The final mass of the remnant (baryonic) was 1.55 M\sol, because
a total of 0.151 M\sol \ of $^{56}$Ni was ejected,
but this was reduced to 0.073 M\sol \ by removing inner zones. 
This amount
of $^{56}$Ni was found necessary by Woosley et al. (1994) to give good
agreement with the observations. Moreover, they found it necessary to
mix the initial chemical composition and $^{56}$Ni in an artificial
fashion. The actual composition of Model~13C, following a
moderate amount of mixing, is shown in Figure \ref{comp}.  This mixed
composition was used in all calculations with EDDINGTON
(see Iwamoto et al. 1997 for the 2D simulations of the mixing in
SN 1993J models). The final
kinetic energy at infinity of the model was $1.2 
\times 10^{51}$ erg.

In STELLA, the presupernova star was constructed in hydrostatic
equilibrium using the pressure-density relation obtained from the same
initially unmixed Model 13C. Then the composition was changed (without
altering the density profile) to the mixed version shown in Figure
\ref{comp} and the model was exploded, not by a piston, but by the
deposition of energy in a layer of mass $\sim0.03$ M\sol \ outside of
1.30 M\sol. After all fell back, the remnant mass was 1.55
M\sol. Since STELLA does not yet include nuclear burning, preservation of the
same mixed composition in the ejecta is assured.

\begin{figure}
\plotone{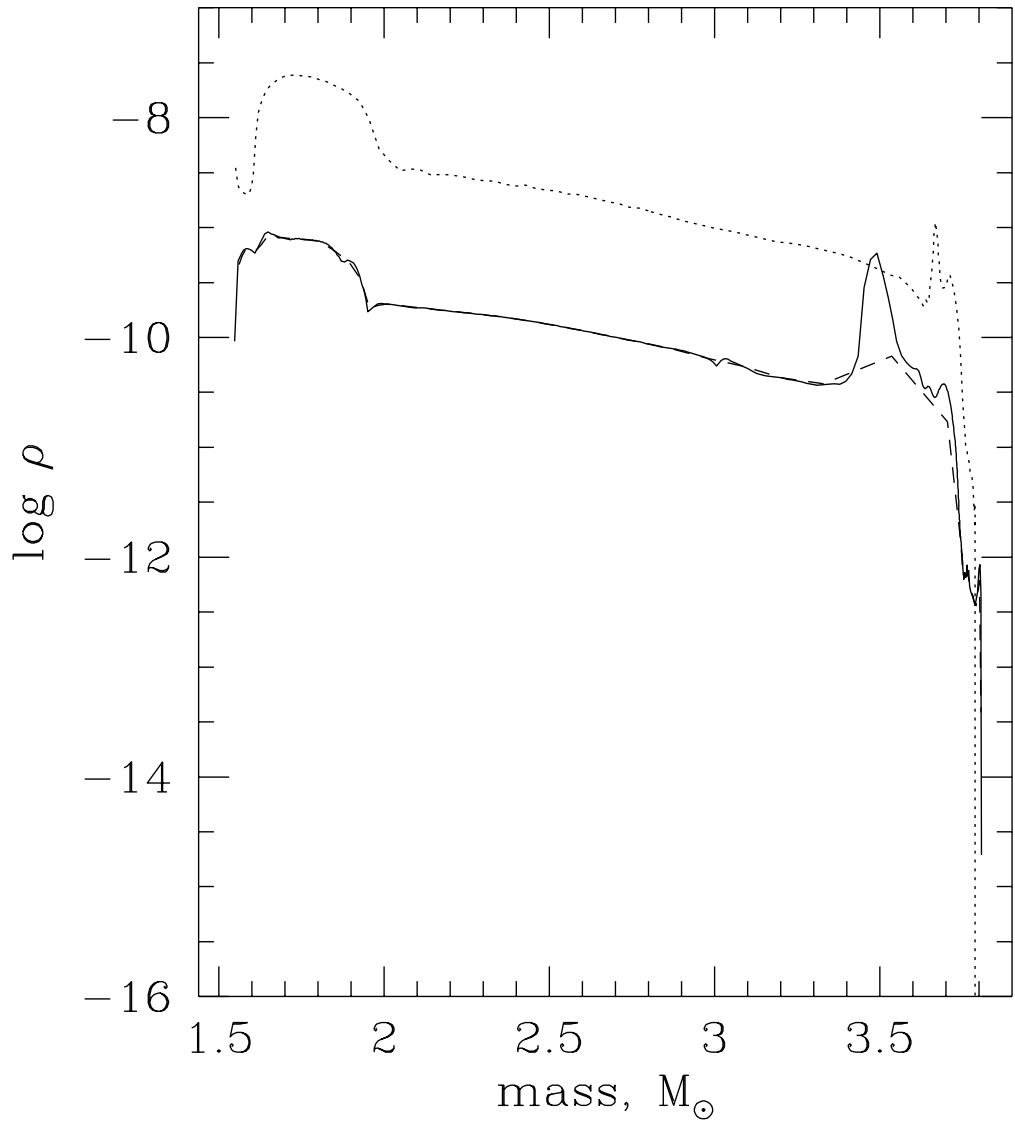}
\caption
{ Density as a function of interior mass for Model 13C 
          during the coasting phase: dotted STELLA ($t=0.78\times 10^5$), 
          solid KEPLER, dashed EDDINGTON (both for $t=2.05\times 10^5$ s)}
\label{prfdm13c}
\end{figure}

\begin{figure}
\plotone{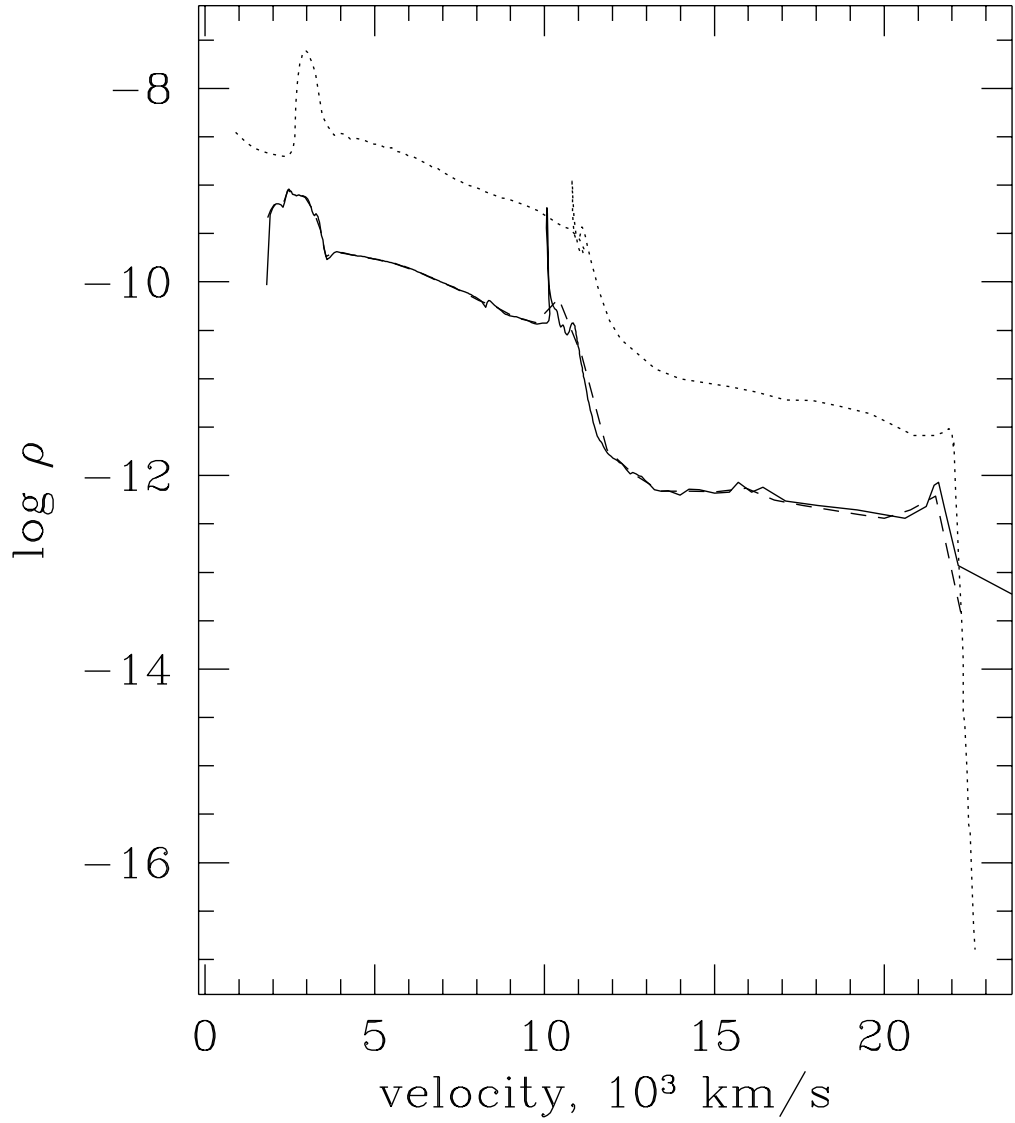}
\caption{Density as a function of velocity for Model 13C 
          at coasting phase: solid STELLA ($t=0.78\times 10^5$), 
          solid KEPLER, dashed EDDINGTON (both for $t=2.05\times 10^5$ s)}
\label{prfdv13c}
\end{figure}

\begin{figure}
\plotone{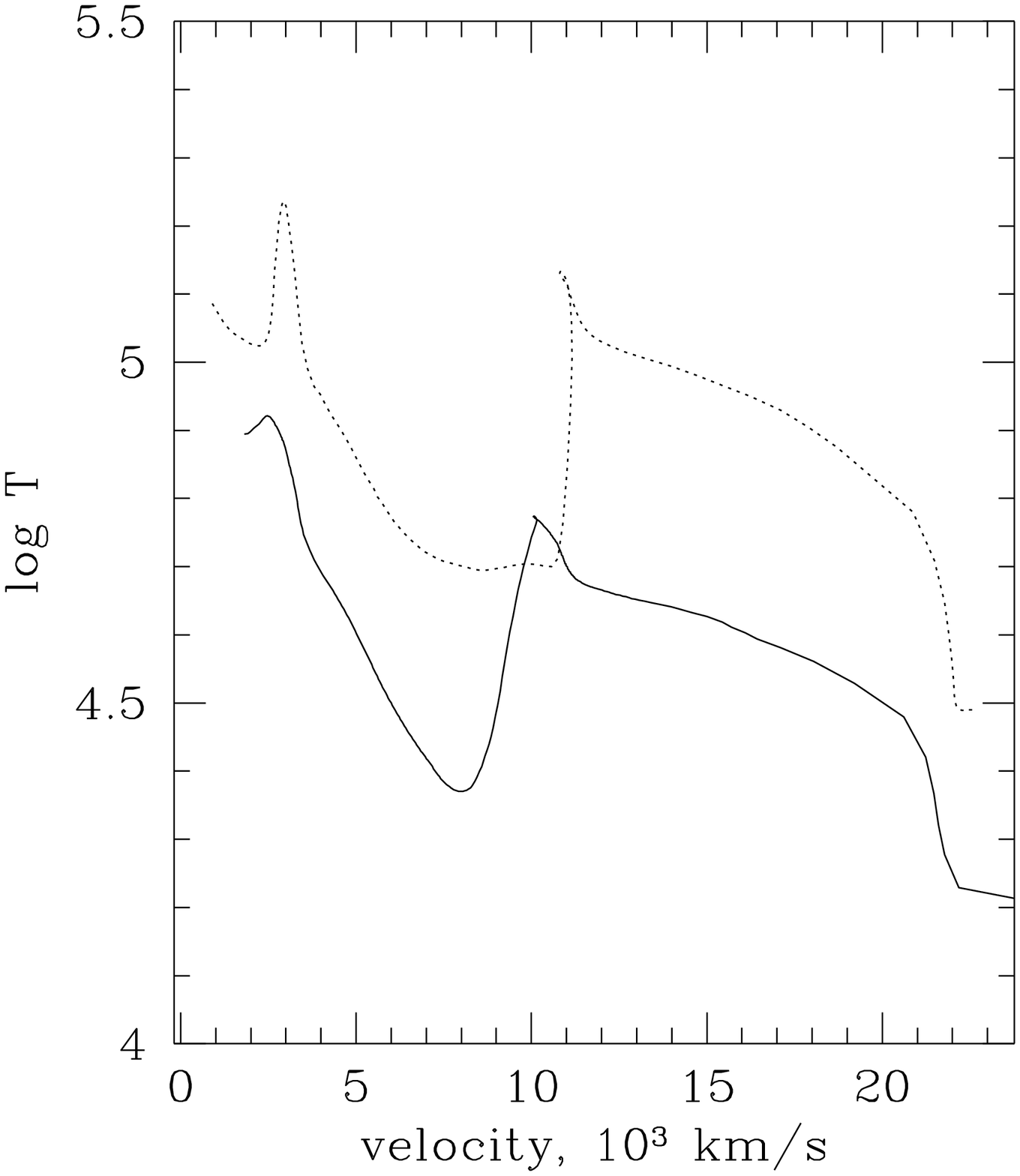}
\caption{Temperature as a function of velocity for Model 13C 
          at early coasting phase: dotted STELLA ($t=0.78\times 10^5$), 
          solid KEPLER ($t=2.05\times 10^5$ s)}
\label{prftv13c}
\end{figure}

The kinetic energy at infinity of the ejecta was $\sim 1.2 \times
10^{51}$ erg in both the KEPLER and STELLA models. 
About $3 \times 10^{50}$ erg of the initially deposited $1.5 \times
10^{51}$ erg  is used to
overcome the gravitational pull of the central core, a small fraction is
radiated away. In spite of the large
differences in mass zoning, mode of simulating the explosion,
and hydrodynamic algorithms, good agreement is found in the
density and temperature profiles (see Fig.~\ref{prfdm13c},
\ref{prfdv13c}, \ref{prftv13c}) at the beginning of coasting phase
(when $v \propto r$).  The EDDINGTON profiles are simply KEPLER ones remapped
from 300 to 80 mesh zones, so the good agreement is not surprising,
but STELLA actually calculates and reproduces even the fine structure
found by KEPLER quite successfully. The dense shell located at
$\sim 3.5$ M\sol \ in the KEPLER model is a a little farther out in
the STELLA one. The formation of this dense shell is caused by a
complex interplay of geometrical factors (first growing $\rho r^3$ and
slowing of the shock, then a drop in $\rho$ and acceleration of the matter)
and the loss of radiation in the outer layers. Since the shock was
dominated and driven by radiation deep inside, the loss of photons
leads to its dramatic slowing down and to formation of the reverse
shock (see Fig. 6 in Woosley et al. 1994). The details of the radiative
transfer are quite different in KEPLER and STELLA, so it is no
surprise that the details of the dense shell do not
coincide. Nevertheless, the qualitative picture is the same.  Note,
that EDDINGTON which preserves a fine zoning in the center of the
model (in order to reproduce accurately the second maximum light of SN
1993J) has a crude zoning here, so the dense shell is artificially smeared.


The similarity of two solutions is better reflected in
Figures \ref{prfdv13c} and \ref{prftv13c} which show good agreement
for the profiles of $\rho$ and $T$ as functions of the velocity, $u$. At
those times $u$ is already $\propto r$ so the radial dependencies are
well represented, with very narrow dense shells. Since $u$ of
each Lagrangian mass is conserved one can directly compare
solutions, which for ideal coincidence must be shifted vertically to a
constant corresponding to the homologous transformation.  A few tenths
of M\sol \ in the outer layers are accelerated to high speed, $ 10 - 20
\times 10^3$ km/s and both hydrodynamic codes reproduce this quite
consistently.

Another set of runs 13C4 -- 13C8 in Table \ref{runs} used as a starting
point just the EDDINGTON remap of the KEPLER model with 80 mesh zones.
The composition (and density) remaps were done in the EDDINGTON
calculations to reduce the computational expense.  The original KEPLER
model had 323 zones, and in EDDINGTON the calculation execution time
scales linearly with the number of zones. The models 13C1-3 used by
STELLA had 200 zones and were very close to the KEPLER one. EDDINGTON
used only 80 zones but employed very fine zoning in the center in order to
reproduce the second maximum light with the best resolution.

\begin{figure}
\plotone{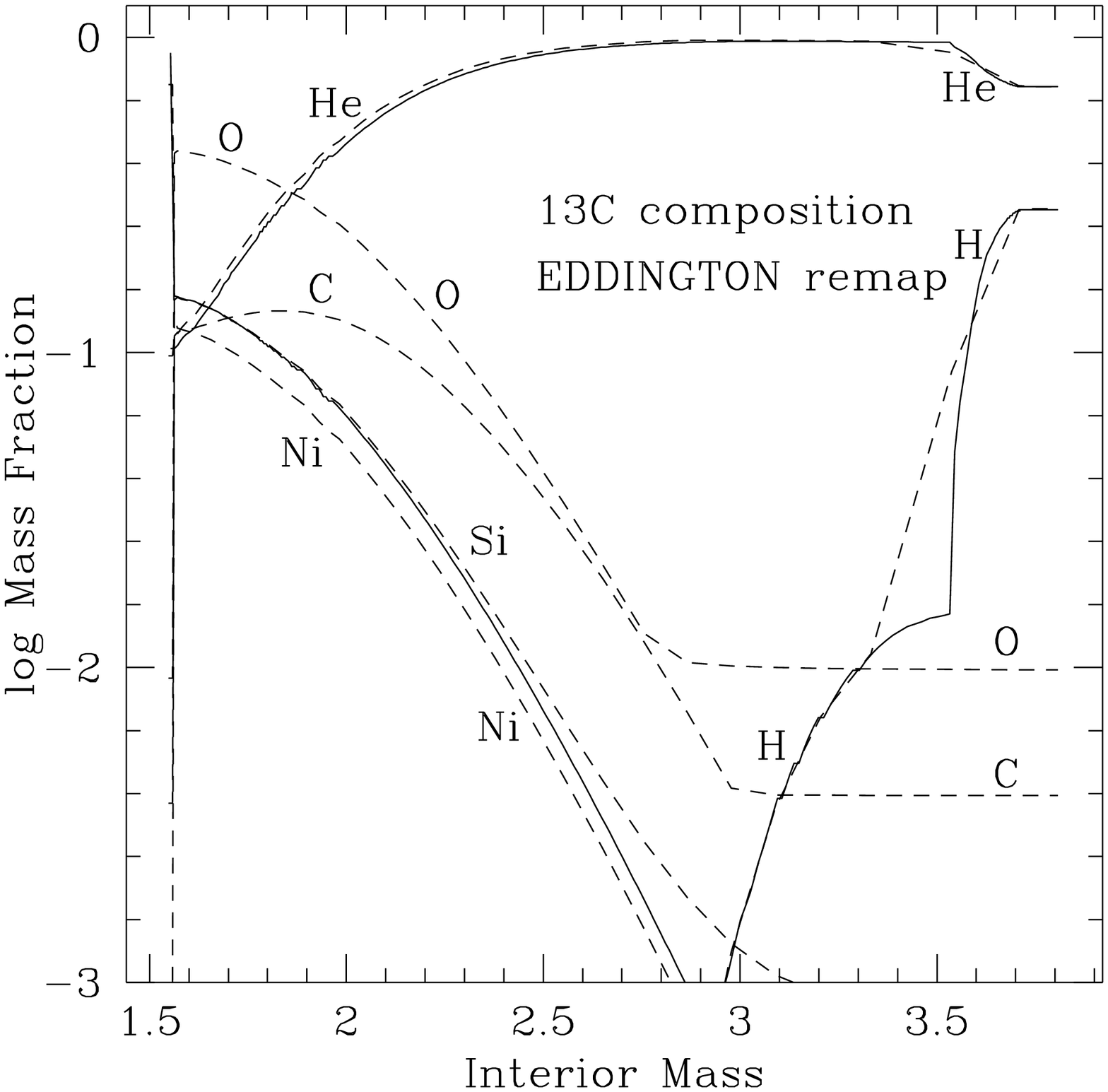}
\caption{The composition in Model 13C as remapped by
 EDDINGTON (dashed). 
Solid lines show original 
 H, He and $^{56}$Ni distribution in 13C}
\label{compRonc}
\end{figure}

\section{THE LIGHT CURVE AT VARIOUS TIMES}

\subsection{General features of the model light curves}

\begin{figure}
\plotone{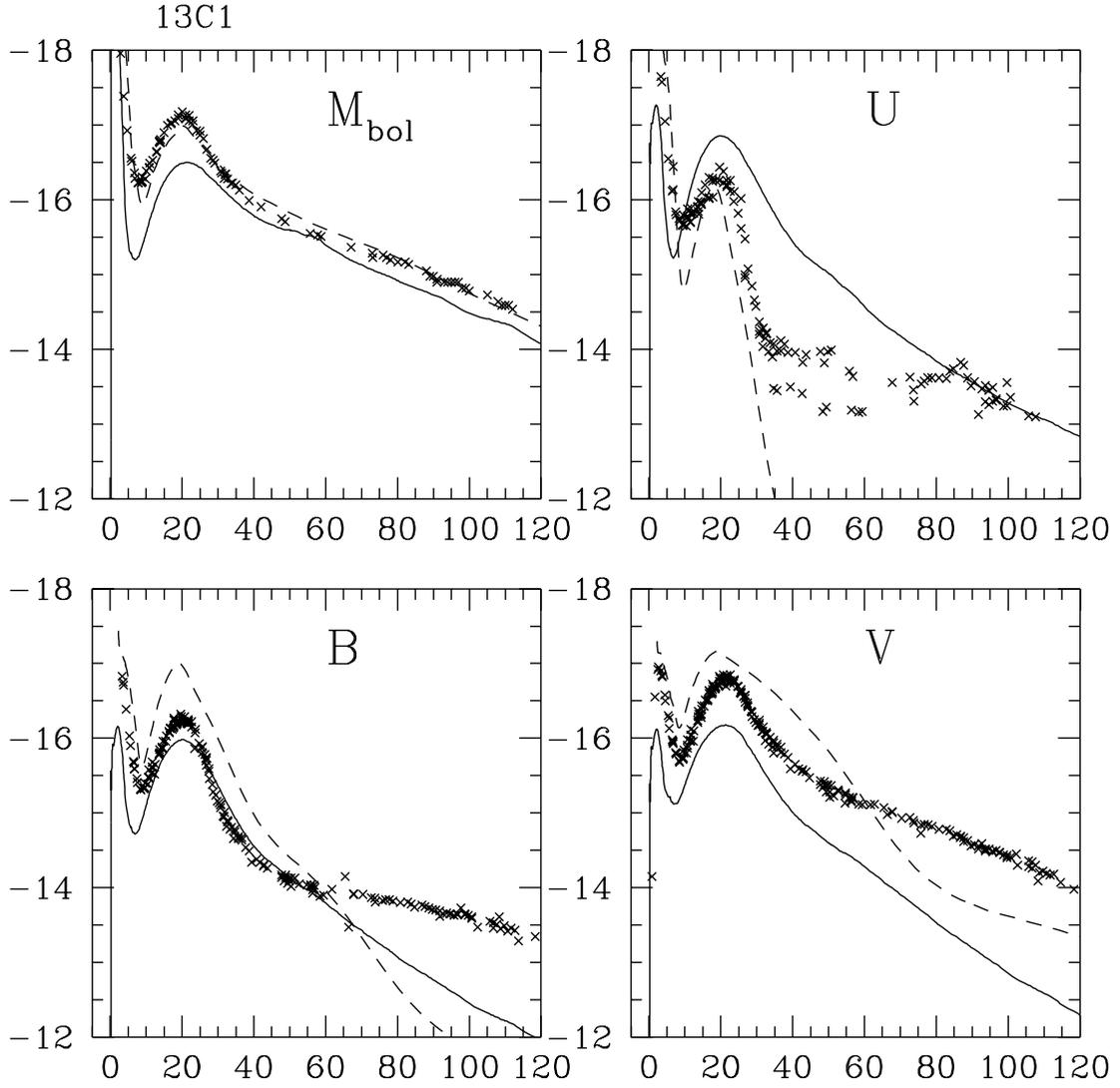}
\caption{ $M_{\rm bol}$ and $UBV$ for Model 13C1 (with scattering dominating in the
 opacity). Results of EDDINGTON are shown by dashed lines
}
\label{mubvs}
\end{figure}

The first runs of STELLA for Model 13C assumed that the line opacity
was scattering dominated (some weak absorption was present according
to Anderson, 1989) and that the effect of expansion small. The
expansion parameter, $t_s$, was set to 100 days.  The resulting curves
for Model 13C1 are the solid lines in Figure\ref{mubvs}.  Dashed lines
are the results of EDDINGTON.  Observations of Richmond et al. (1994)
are shown by crosses.  We see that Model 13C1 produces, for this form
of opacity, too little luminosity both in comparison with
observations and with EDDINGTON.  $UBV$ fluxes also
indicate that the model is too hot (too bright in $U$).
Figure\ref{mubvs} gives all values in the frame comoving with the outer
radius of the supernova model (below we discuss the transformation to
the rest observer frame), but it is already clear that this correction,
which is of order $u/c$ for $M_{\rm bol}$, cannot explain such a large
discrepancy. The effect of the expansion on opacity is also 
insufficient (see below).

The chief difference here is the treatment of the line
opacity. In the EDDINGTON calculation, the line opacity was treated as 
purely absorptive (possible justification for this is discussed below). Near
the secondary maximum ($t \simeq 20$ days) the models are already
semi-transparent and the emission, which is low for the case of
scattering dominated line opacity, can be enhanced by {\em increasing}
the absorptive opacity (while keeping the total extinction the
same). This is a consequence of the Kirchhof's law,
$\bar\eta_\nu=\chi_{\rm a} b_\nu(T)$ for the emission coefficient
$\bar\eta_\nu$ in (\ref{comov}).

\begin{figure}
\plotone{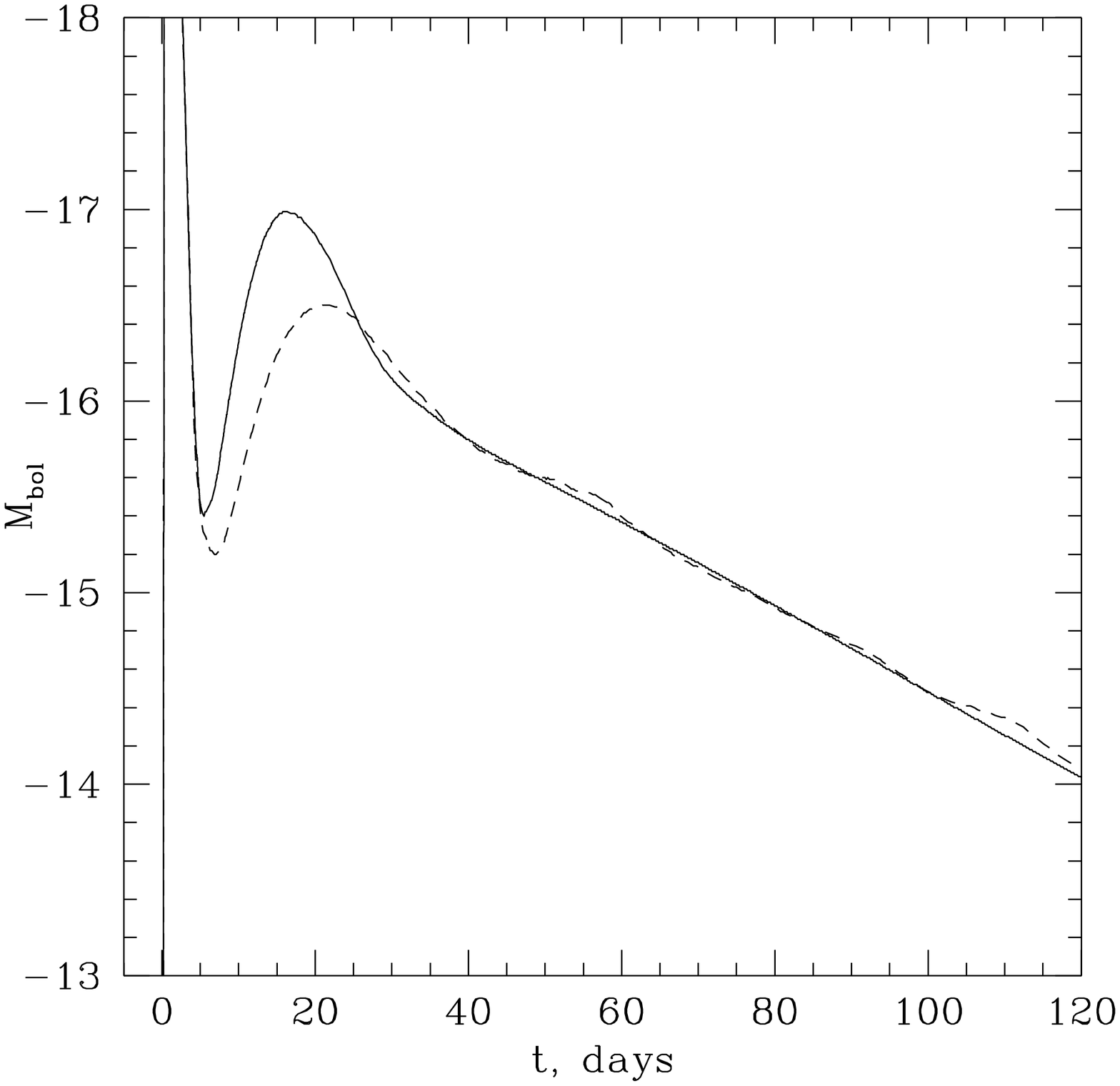}
\caption{Comparison of $M_{\rm bol}$  
 for run 13C1 with scattering lines (dashed) and 13C2 with 
 forced absorption (solid)}
\label{mbolas}
\end{figure}

In order to maximize the effect we assumed, in calculation 13C2, that
{\em all} the extinction was purely absorptive.  Figure \ref{mbolas}
compares directly the two bolometric light curves of 13C1 (scattering)
and 13C2 (absorptive). The difference is striking.  In the {\it
scattering} case (dashed line) the brightness at the second peak is
half a magnitude fainter. The total opacity (the extinction) was the
same in both calculations, and one cannot say that the effective
opacity is larger in the case were the line opacity is taken as
scattering. Such a result is indicative of the fact that by the time
of the second peak, the radiation field is not in the equilibrium
radiative diffusion (ERD) regime, where $J_\nu\approx B_\nu(T)$ and
$H_\nu\approx -(1/\chi_\nu)(\partial B_\nu\bigl(T(r)\bigr)/\partial
r)$ and $\chi_\nu$ is the total opacity. In ERD it makes no difference
whether the opacity is absorptive or scattering, but here one can see
directly the effect of enhanced volume emission due to the enhanced
absorption.

The assumption of full absorption in the run 13C2 may overestimate
the effect, but we shall see that if one assumes the opacity
is purely absorptive only in lines, the difference with this
maximum absorption case is not large. Lines dominate the total
extinction in the case under consideration.

What is the physical justification for ascribing this enhanced
absorption to a process that is, microscopically, actually scattering?
When a photon Doppler shifts into resonance with a line and is
absorbed, the quanta will scatter around in the line until one of 
three things happens: the photon may escape from the resonance region,
in which case we would regard the interaction as a scattering event.
The excited state could be depopulated by an electron collision, in
which case we would regard the interaction as an absorption event. A
third possibility is that the excited state is depopulated radiatively
via an alternate channel, and one or more fluorescence photons
emitted. This last possibility turns out to be important.

Considering only the upper and lower levels of a transition, the
probability that a trapped photon is collisionally depopulated may be
written
\begin{eqnarray}
p(\hbox{therm})={C_{ul}\over C_{ul}+
A_{ul}\beta_{ul}}\approx\nonumber\\
\approx {N_e\over N_e + q \beta_{ul}\left({h\nu / 1\ \hbox{eV}}\right)^3}
\label{thermprob}
\end{eqnarray}
where $C_{ul}$ is the downward collision rate, $A_{ul}$ is the
spontaneous decay rate, and $\beta_{ul}$ is the Sobolev escape
probability, given by $\beta_{ul}=(1-\exp(-\tau_{ul}))/\tau_{ul}$. The
last expression on the right is from Anderson (1989) who, by using the
Van Regemorter (1962) approximation and oscillator strengths from
Kurucz finds that for iron, $q\sim 10^{13}$ to $10^{14}\
\cc$.  In Model~13C at 15~days, $N_e\ltaprx10^{11}\ \cc$. Taking
$q=10^{13}\
\cc$ and $h\nu=6$~eV, we get the result that only lines with
\begin{equation}
\tau_{ul}\gtaprx 2\times10^4 \left({10^{11}\ \cc\over N_e}\right)
\end{equation}
are likely to experience collisional destruction. Much of the UV line
opacity however comes from transitions with $\tau_{ul}\ltaprx 1$. One
might be tempted to conclude that the line opacity should be treated
as scattering. However this analysis neglects another possibility,
that of fluorescence. Pinto and Eastman (1997) analyzed the relative
probabilities of escape, thermal destruction and fluorescence and
found, for similar physical conditions, that fluorescence may be as
much as 5 to 10 times more likely than escape. This mechanism allows a
short wavelength photon to be split into two or more long wavelength
photons which see a smaller opacity and optical depth to the surface.
Within the context of LTE, a crude way to take this process into
account is by assuming that photons absorbed in line transitions are
thermally destroyed. This provides a mechanism for them to be
thermally re-radiated at longer wavelengths, but it is only a crude
representation of the real physics, which involves simulating the
fully non-LTE population kinetics.  Nugent et al [1995,1997] treated
many lines of Fe~II in full non-LTE, but found that, in order to match
observations, they had to introduce a large value of the
thermalization parameter for other lines in their models.  As pointed
out above, the "naive" assumption that lines should be treated as pure
scattering seems pretty reasonable if one asks the question, "given
the relevant range of electron densities, what is the probability that
collisional deexcitation will thermalize a photon in resonance with a
line". Treating the line opacity as purely absorptive is a crude,
zeroth order attempt at accounting for the effect of splitting in an
LTE calculation.  See also, for example, H\"oflich (1995), and Li\&
McCray (1996) on other approaches to the fluorescence effect, and
Jefferies (1968) and Canfield (1971) for early attempts to account for
it.

\begin{figure}
\plotone{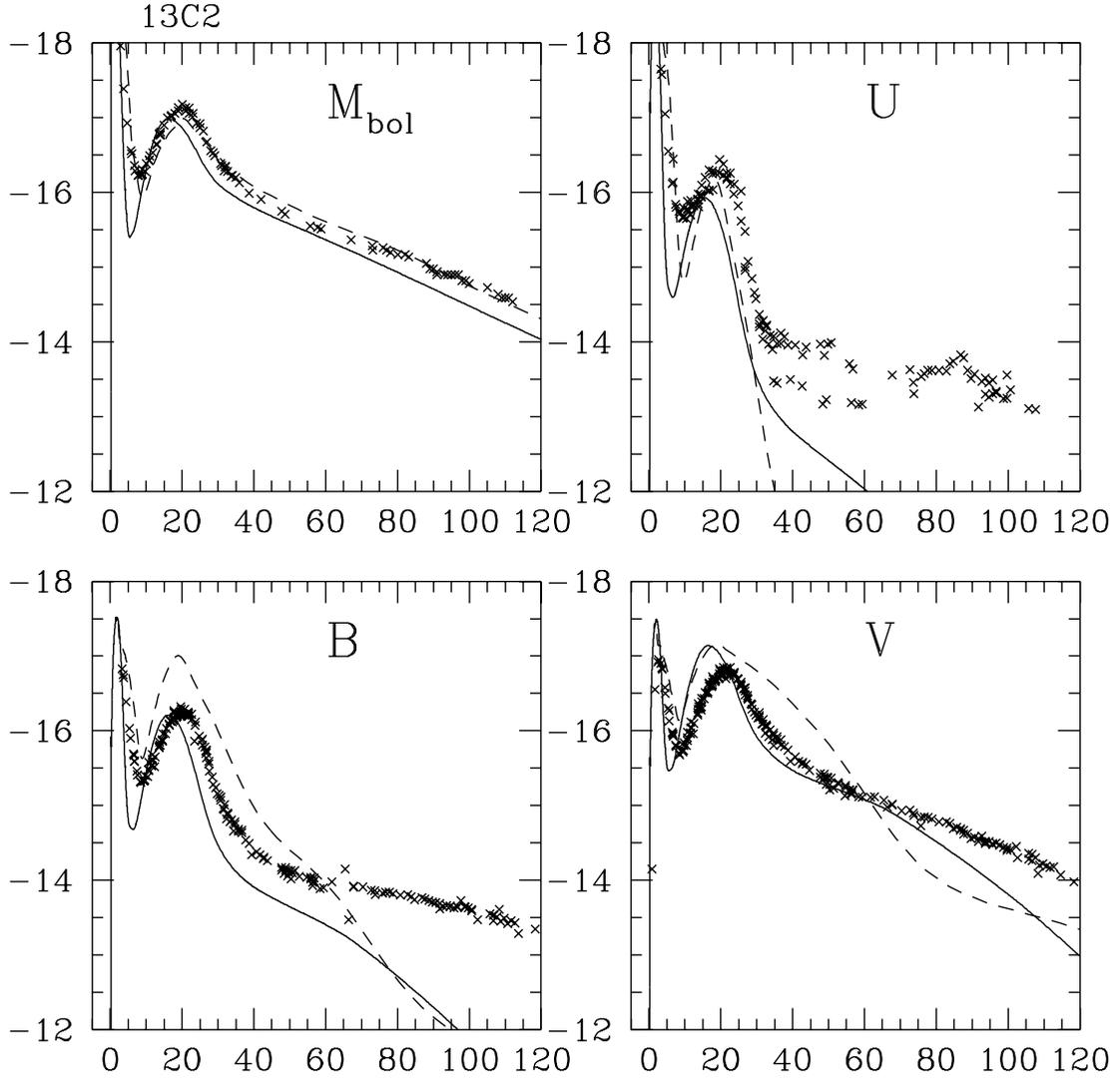}
\caption{ $M_{\rm bol}$ and $UBV$ for the run 13C2 
   (forced absorption) shown by solid lines. EDDINGTON - dashed lines}
\label{mubva}
\end{figure}

Figure \ref{mubva} gives the resulting light curves for the forced
absorption case of Model 13C2.  
With enhanced absorption, the STELLA results are in much better
agreement with those of EDDINGTON, and with the observations of
SN~1993J.
At this level of accuracy it becomes worthwhile to take into
account the corrections for the transformation to the observer frame.

\vskip 1 true cm

\subsection{Transforming to the observer frame}

EDDINGTON solves for both the comoving
frame intensity and the low order angular moments (energy
density and flux), at each time step, so it is straightforward to
transform the angle dependent specific intensity into the
frame of an observer at rest to obtain the observer frame flux.
In STELLA however, the specific intensity is not updated at every time
step, although the comoving frame energy density and fluxes are.
To transform the comoving frame surface flux into the observer frame,
one could use the formula proposed by 
Mihalas \& Mihalas (1984, eq.~[99.39])
\begin{equation}
   F_\nu(observer)=F_\nu+u(E_\nu + P_\nu)
                      -[\partial(\nu u P_\nu)/\partial\nu ] \; .
\label{FnuXform}
\end{equation} 
However, this formula assumes that the $P_\nu$ is
smoothly varying with frequency, whereas there are usually large
jumps in flux with frequency.  Another approach was necessary in
STELLA.

To find the observer frame flux, $F^o_{\nu_o}$, the following procedure is
performed: the radially propagating photon received by an observer at
rest, at frequency $\nu_o$, has the Doppler-shifted frequency
$\nu=\nu_o(1-u/c)$ in the comoving frame at the outer radius, where
$u$ is the speed of the outermost layers.  The idea is to assume that
the flux for the observer at rest, $F^o_{\nu_o}$, is close to the
comoving flux, $F_\nu$.  The frequency $\nu$ falls into an interval
$(\nu_j, \nu_{j+1})$ of the frequency grid in the comoving frame. 
$F^o_{\nu_o}$ is logarithmically interpolated between
the comoving frame fluxes $F_{\nu_j}$ and $F_{\nu_{j+1}}$: 
\begin{equation}
  F^o_{\nu_o} = C_F(\nu_o) \exp[w\ln F_{\nu_j} 
                   + (1-w) \ln F_{\nu_{j+1}}] \; .
\label{fluob}
\end{equation} 
The weight $w$ is simply the relative
distance of the frequency $\nu$ from the grid points on a logarithmic
scale: $w=\ln(\nu_{j+1}/\nu)/\ln(\nu_{j+1}/\nu_j)$, and $C_F(\nu_o)$
is a correction coefficient: it measures how close $F^o_{\nu_o}$ is to
$F_\nu$.  This correction factor is found from solution of the
transport equation for the specific intensity $I_\nu(\mu)$, (which we
reiterate, is not performed at every time step).  The integration
yields $I_\nu(\mu)$ for a set of angles with cosines $\mu$, which is
Lorentz transformed into the observer frame and used to compute the
observer frame flux, $\tilde{F}^o_{\nu_o)}$. From $I_\nu(\mu)$ a
comoving frame flux is also computed, $\tilde{F}_{\nu}$. One then
writes
\begin{equation}
\tilde{F}^o_{\nu_o} = C_F(\nu_o) \exp[w\ln \tilde{F}_{\nu_j} 
                       +  (1-w) \ln \tilde{F}_{\nu_{j+1}}] \; ,
\label{fluos}
\end{equation}
which, since $\tilde{F}^o_{\nu_o}$ and $\tilde{F}_{\nu}$ are known,
gives the correction coefficient $C_F(\nu_o)$.  We find that
$C_F(\nu_o)$ changes only slowly between time steps, so we get the
observed fluxes $F_{\nu_o}$ using values of $C_F(\nu_o)$ determined at
the last time step at which the specific intensity and Eddington
factors were updated.  To preserve global energy balance, the
monochromatic fluxes are normalized to the observer frame bolometric
flux, $F^o$, which is easily obtained from frequency integrated
quantities in the comoving frame using
\begin{equation}
     F^o = F + u(E + P)\ ,
\label{flutob}
\end{equation}
which may be obtained by integrating equation \ref{FnuXform} over all $\nu$.

\begin{figure}
\plotone{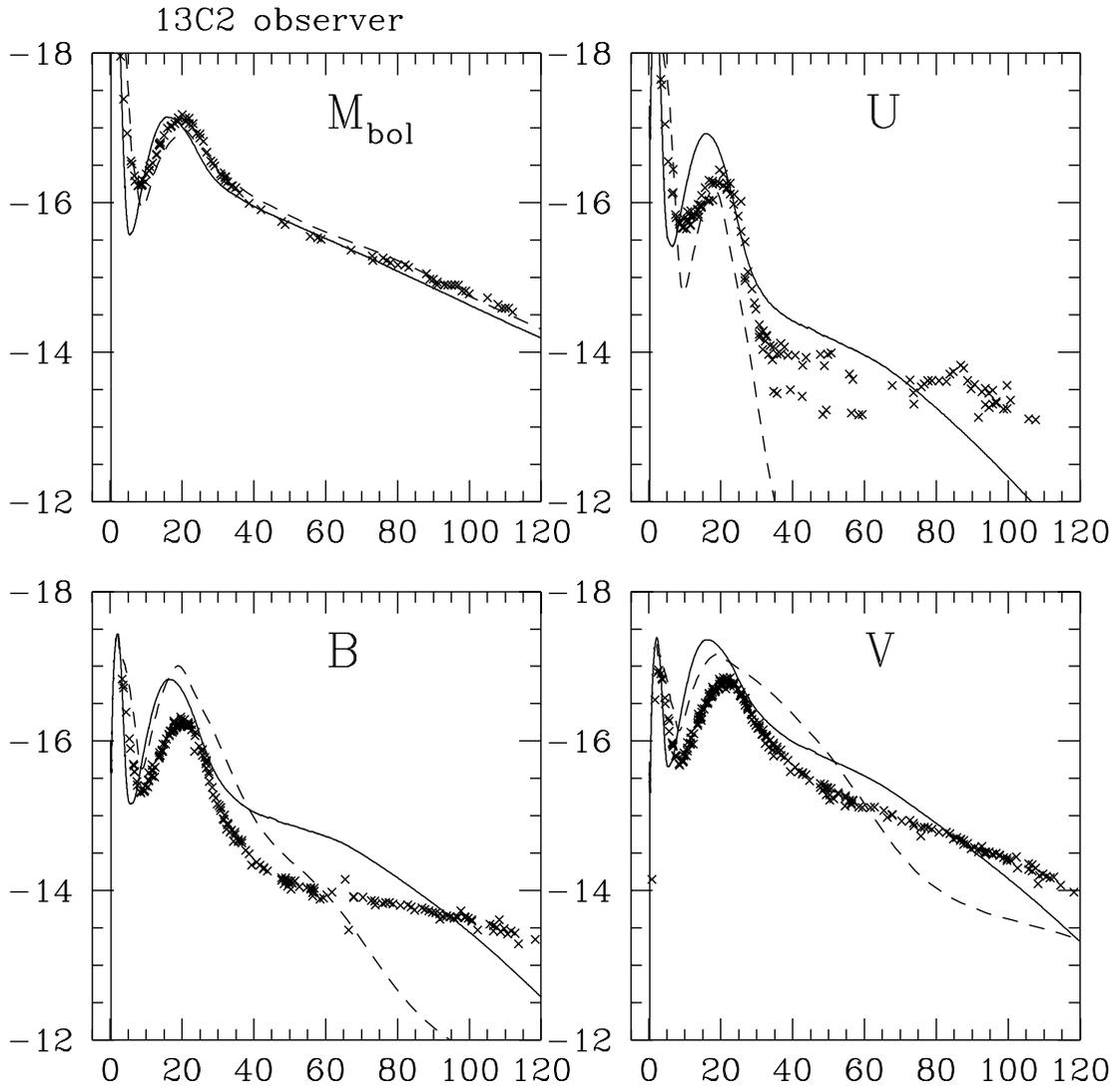}
\caption{The same as in Fig \protect\ref{mubva} but STELLA results
   (solid lines) are now transformed to the observer frame}
\label{13C2ob}
\end{figure}

One can compare the result of the transformation for Model 13C2 in
Figure \ref{13C2ob} with the same model in the comoving frame in
Figure \ref{mubva}. The effect of the transformation is larger for the
broad band fluxes than for $M_{\rm bol}$, especially for $U$, since
now at a given wavelength one observes photons which were at shorter
wavelengths in the comoving frame, and in the blue part of the
spectrum the continuum flux $\lambda$ variation is larger,
approximately obeying the Wien's law.

In both cases we see that the minimum in $M_{\rm bol}$ for $t\sim 7 -
10$ days produced by STELLA is more pronounced than that of
EDDINGTON. Tests showed that this result is very sensitive to the
zoning of the model and abundance distribution (compare
Figs. \ref{comp} and \ref{compRonc}), so it is more instructive to
compare the results of the two codes for the exactly the same zoning.

\vskip 1 true cm

\subsection{The influence of physical assumptions on the light curves}

\begin{figure}
\plotone{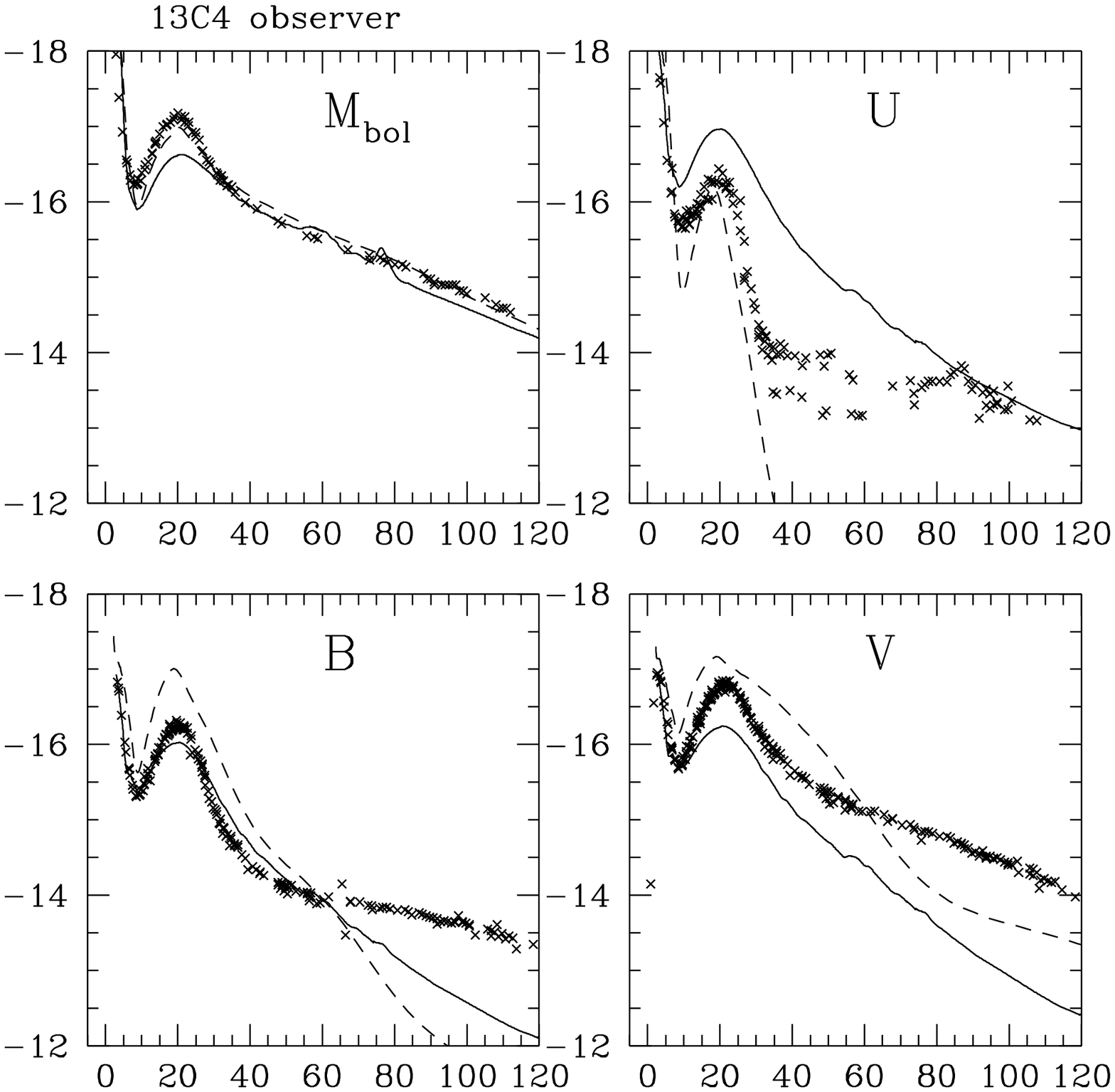}
\caption{ $M_{\rm bol}$ and $UBV$ for the run 13C4 of STELLA 
 (scattering dominated
 line opacity and the same zoning as used by EDDINGTON) 
   shown by solid lines. EDDINGTON - dashed lines}
\label{13C4}
\end{figure}

\begin{figure}
\plotone{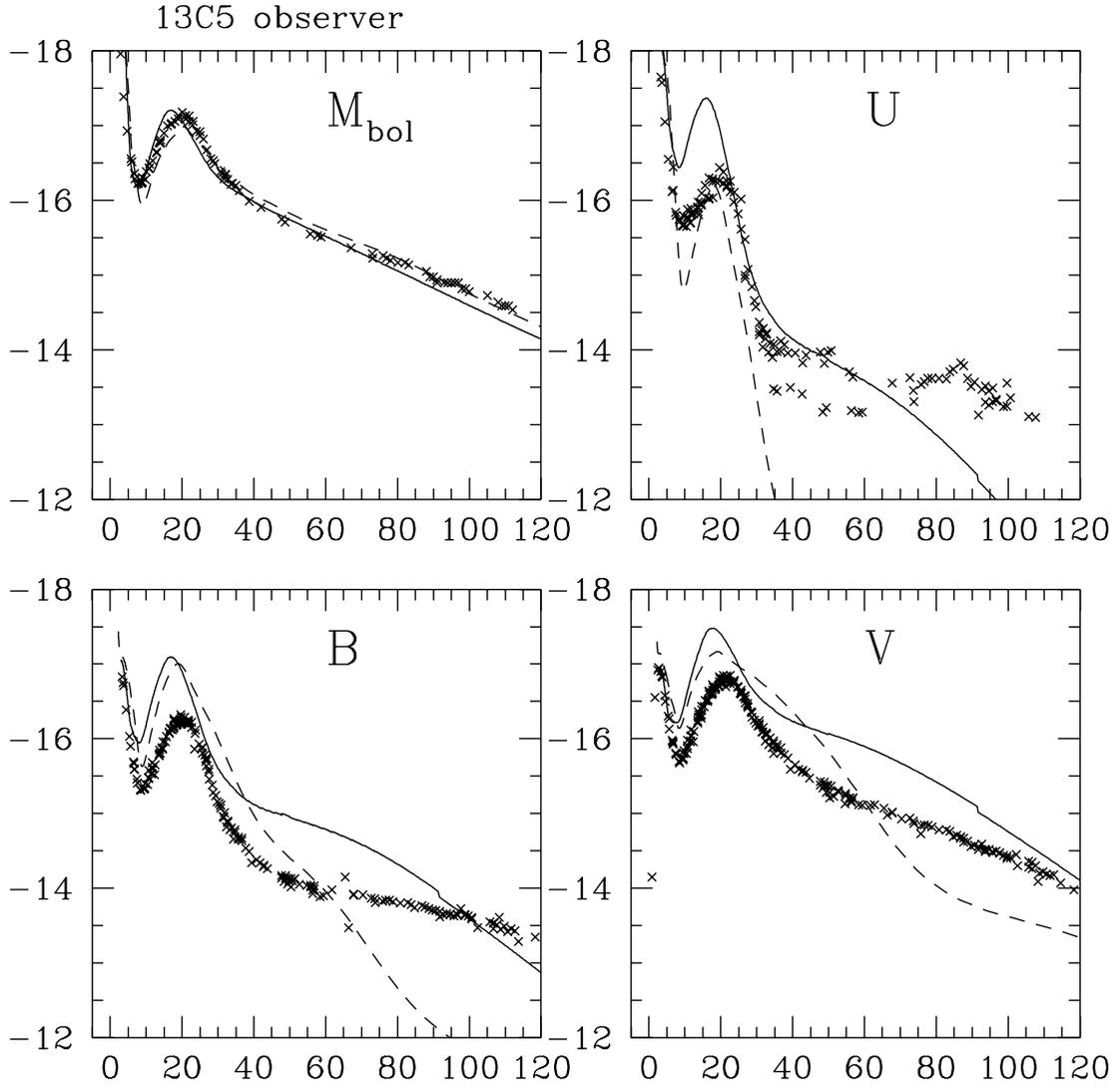}
\caption{ $M_{\rm bol}$ and $UBV$ for the run 13C5
   (absorption dominated lines) shown by solid lines. 
    EDDINGTON - dashed lines}
\label{13C5}
\end{figure}

\begin{figure}
\plotone{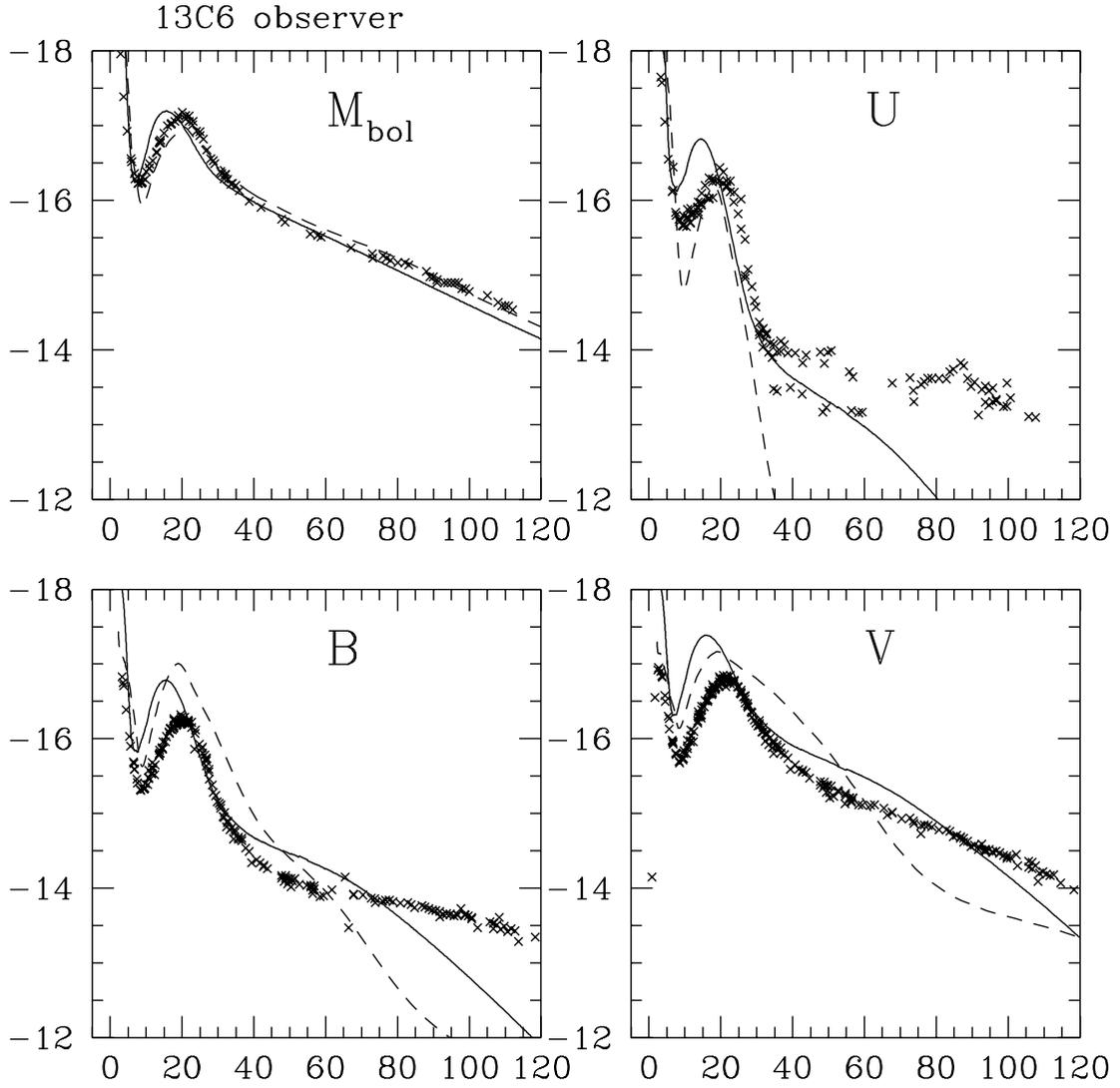}
\caption{ $M_{\rm bol}$ and $UBV$ for the run 13C6
   (forced absorption for all types of extinction) shown by solid lines. 
    EDDINGTON - dashed lines}
\label{13C6}
\end{figure}

\begin{figure}
\plotone{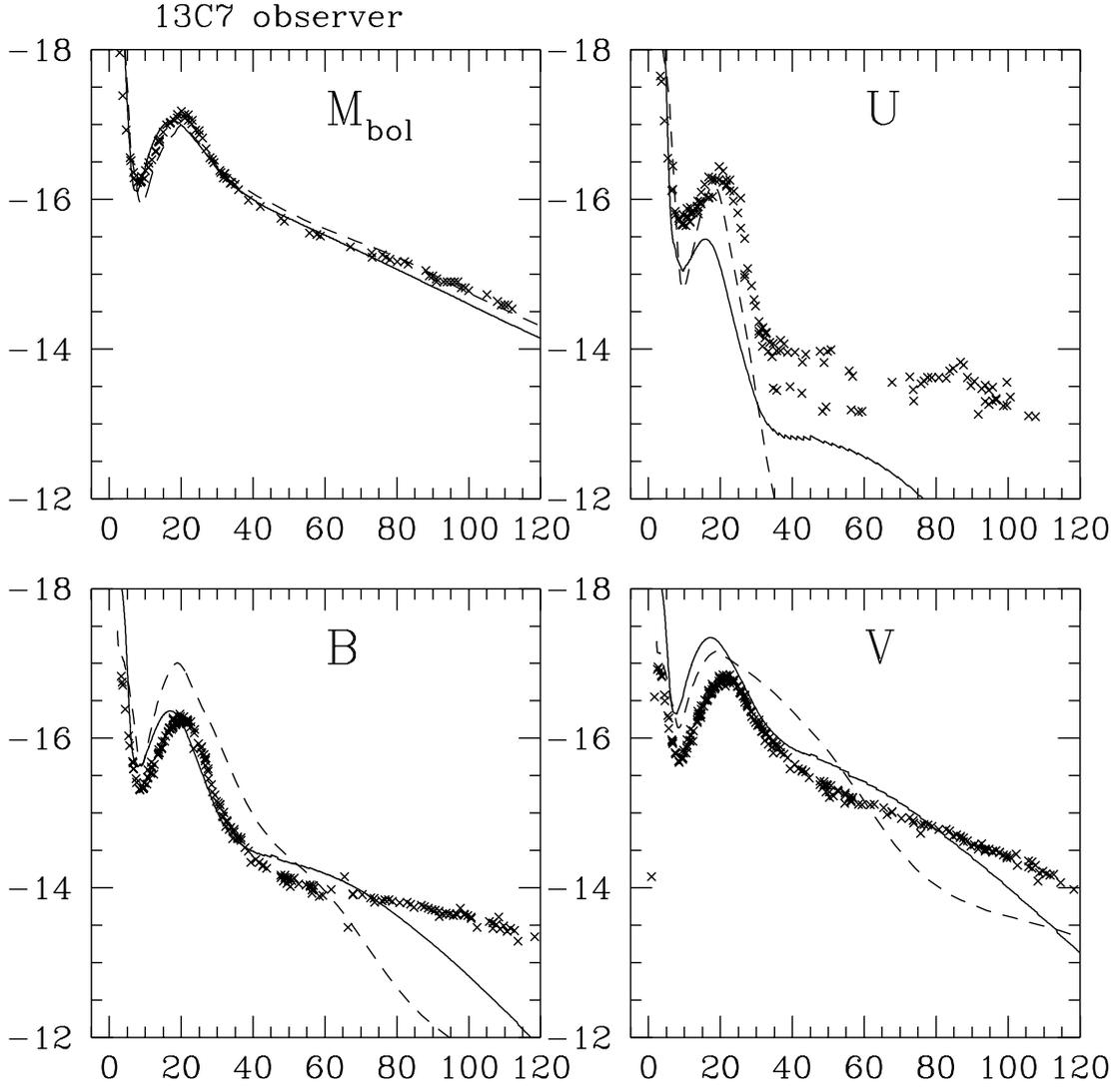}
\caption{ $M_{\rm bol}$ and $UBV$ for the run 13C7
   (forced absorption as in Figure \protect\ref{13C6}
    but now with full expansion effect according to
    Eq.\protect\ref{expopac}) shown by solid lines. 
    EDDINGTON - dashed lines}
\label{13C7}
\end{figure}

\begin{figure}
\plotone{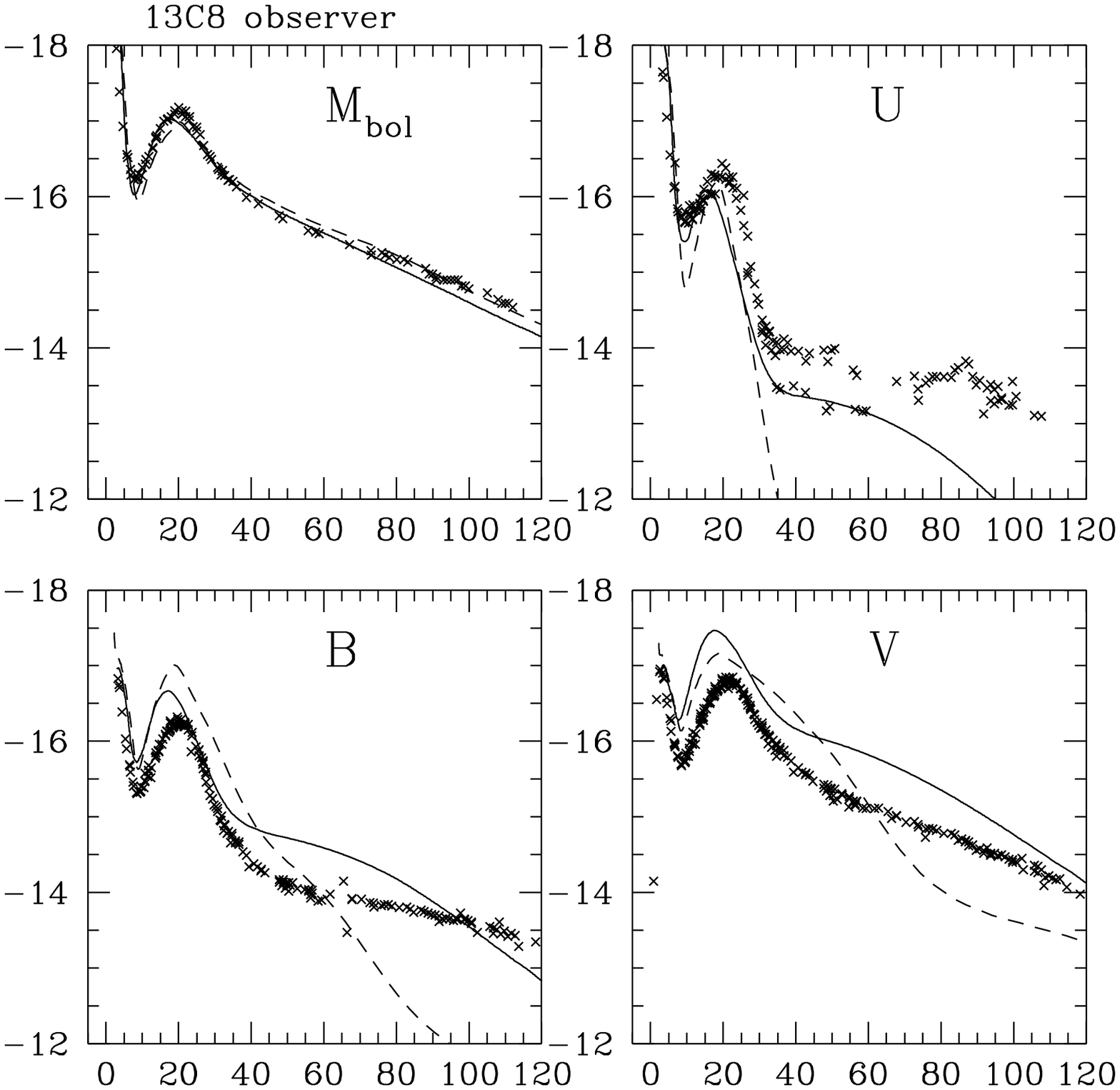}
\caption{ $M_{\rm bol}$ and $UBV$ for the run 13C8
   (absorption dominated lines as in Figure \protect\ref{13C5},
    but now with full expansion effect according to
    equation \protect\ref{expopac}) shown by solid lines. 
    EDDINGTON - dashed lines}
\label{13C8}
\end{figure}

Runs 13C4 -- 13C8 in Table \ref{runs} used exactly the same zoning and chemical
composition as was used by EDDINGTON.  The results are given in
Figures \ref{13C4} -- \ref{13C8}. 


The run  13C4 with scattering dominated lines is similar to 13C1, but
now all fluxes are given in the observer's frame, and are found to be
too low in comparison with observations.

The runs 13C5 (absorption in lines) and 13C6 (forced absorption for
all processes) are very similar. The bolometric light curve is already
very close to the results of EDDINGTON, but the $U$ flux is higher.
The explanation of this last discrepancy is simple: the runs
13C4 -- 13C6 assumed virtually no expansion effect in the line opacity,
the parameter $t_s=100$ days, and the $U$ band is especially sensitive
to the expansion effect of ultraviolet forest of metal lines. So
for the runs 13C7 and 13C8 we included the full expansion effect
according to the expression (\ref{expopac}) and using the following
interpolation algorithm:
for five values of the velocity gradient, measured as $1/t_s$, 
\begin{equation}
 t_s=1  \quad  3.16 \quad  10 \quad 31.6 \quad 100 \quad {\rm days} \; ,
\end{equation}
the opacity for each mass zone was saved in tables.
During the runs 13C7 and 13C8 we have used
the logarithmic interpolation of those opacity tables for current
value of the time after explosion $t$ between the corresponding tabulated
values of $t_s$.

\begin{figure}
\plotone{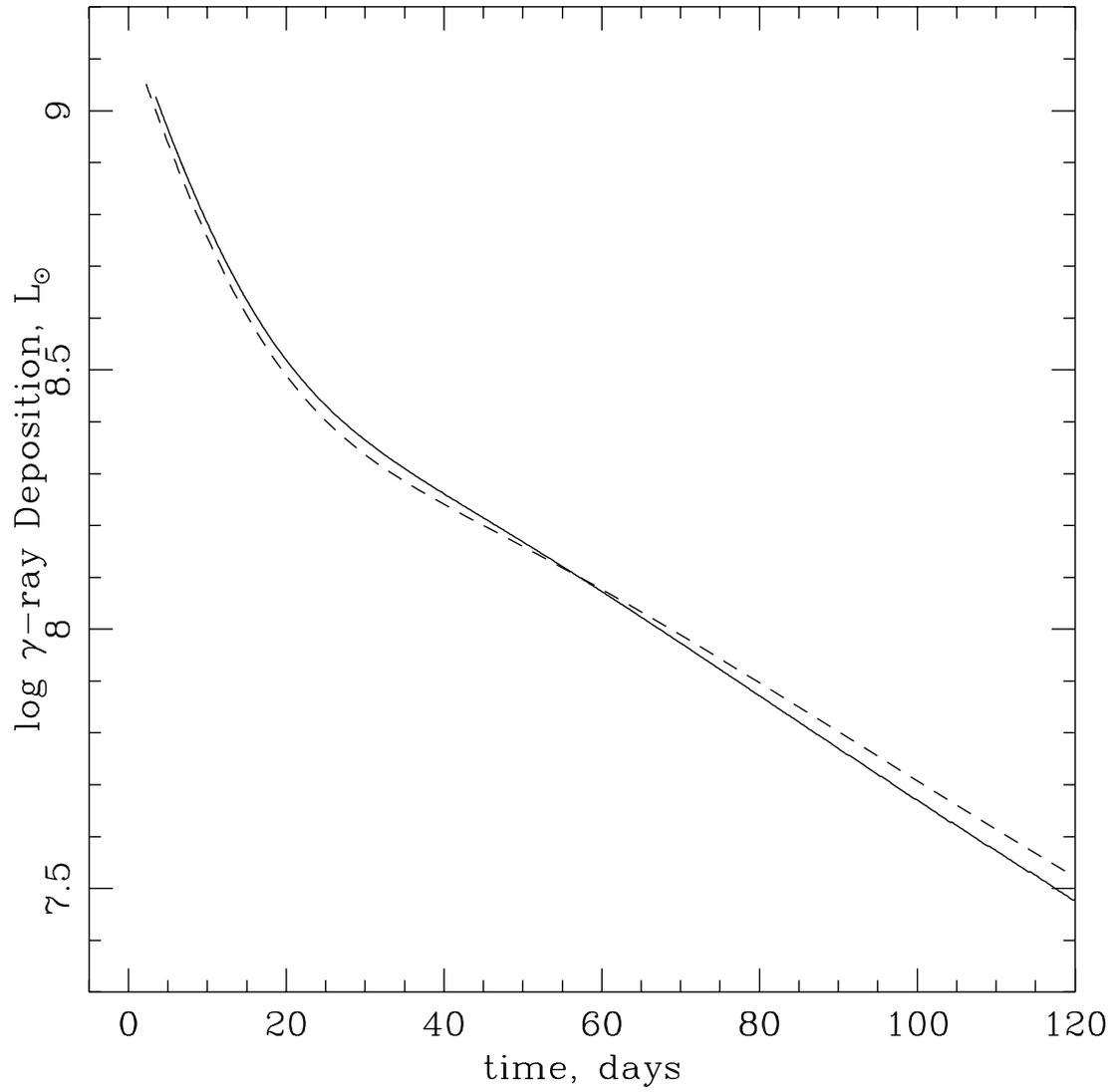}
\caption{ Logarithm of gamma-ray deposition luminosity,  L\sol, 
          in Models 13C4 - 13C8. Solid STELLA, dashed EDDINGTON}
\label{gdeplg}
\end{figure}

The effect of expansion opacity brings the bolometric light curves
very close to the EDDINGTON's. A small deviation on the tail is
completely explained by a slight difference of the gamma-ray
deposition routines.  We compare the energy deposited by gamma-rays
into the matter in the two codes in Figure~\ref{gdeplg}, and one can see
that both codes do the tail of the bolometric light curve consistently.

The effect of expansion is especially strong in the $U$ band, and the
best agreement of two codes is reached, as anticipated, in the run
13C8 where all physics in STELLA follows the physical input of
EDDINGTON most closely.
 
In the test run 13C2s we also checked the effect of a crude frequency
discretization (maximum 20 frequency bins used instead of 100 bins
in the standard STELLA runs). We found that $M_{\rm bol}$ is reproduced quite
well even with this crude frequency grid, but the $UBV$ fluxes are much less
reliable. So full hydrodynamics runs of STELLA with a small number
of frequency bins can be used for fast preliminary calculations of
bolometric light curves, but a detailed comparison with the broad
band photometry requires a reasonably large number of frequency bins.

\subsection{Shock break-out}

Given the good agreement of the STELLA results both with hydrodynamics
of KEPLER and the radiative transfer of EDDINGTON, one trusts that
reliable results might be obtained for the continuum spectra of SN
1993J near shock break-out, a time when neither of the two earlier
codes could operate accurately. For SN 1993J the problem may be more
complicated since the progenitor was distorted by a companion (Woosley
et al. 1994), Still the results should still be approximately correct.

\begin{figure}
\plotone{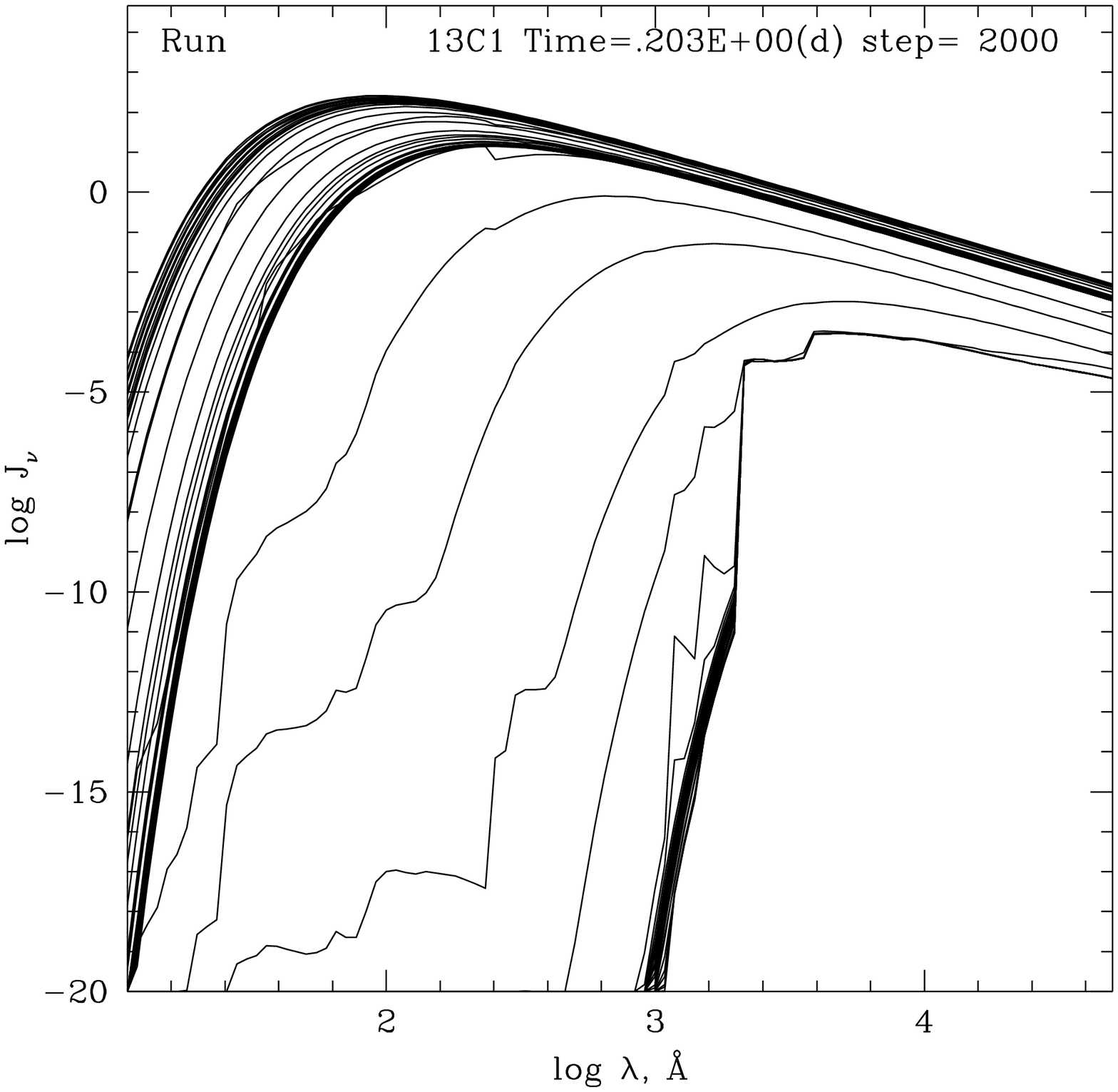}
\caption{Mean comoving-frame intensity $J_\nu$ (in arbitrary units) 
    in all mass zones just prior to  shock breakout in the run 13C1.
    The uppermost curves correspond to the innermost layers
 and the lowest curve to the spectrum on the surface of the star.}
\label{break1}
\end{figure}

Figure \ref{break1} gives the mean intensity distribution found by
STELLA inside the star just prior to shock break-out. In the
intermediate layers the intensity departs strongly from a blackbody
since in the extreme ultraviolet the radiation from the deep layers
already heated by the shock wave becomes visible. This complicated
behavior would be poorly represented in a hydro code that employed only
one energy group.

\begin{figure}
\plotone{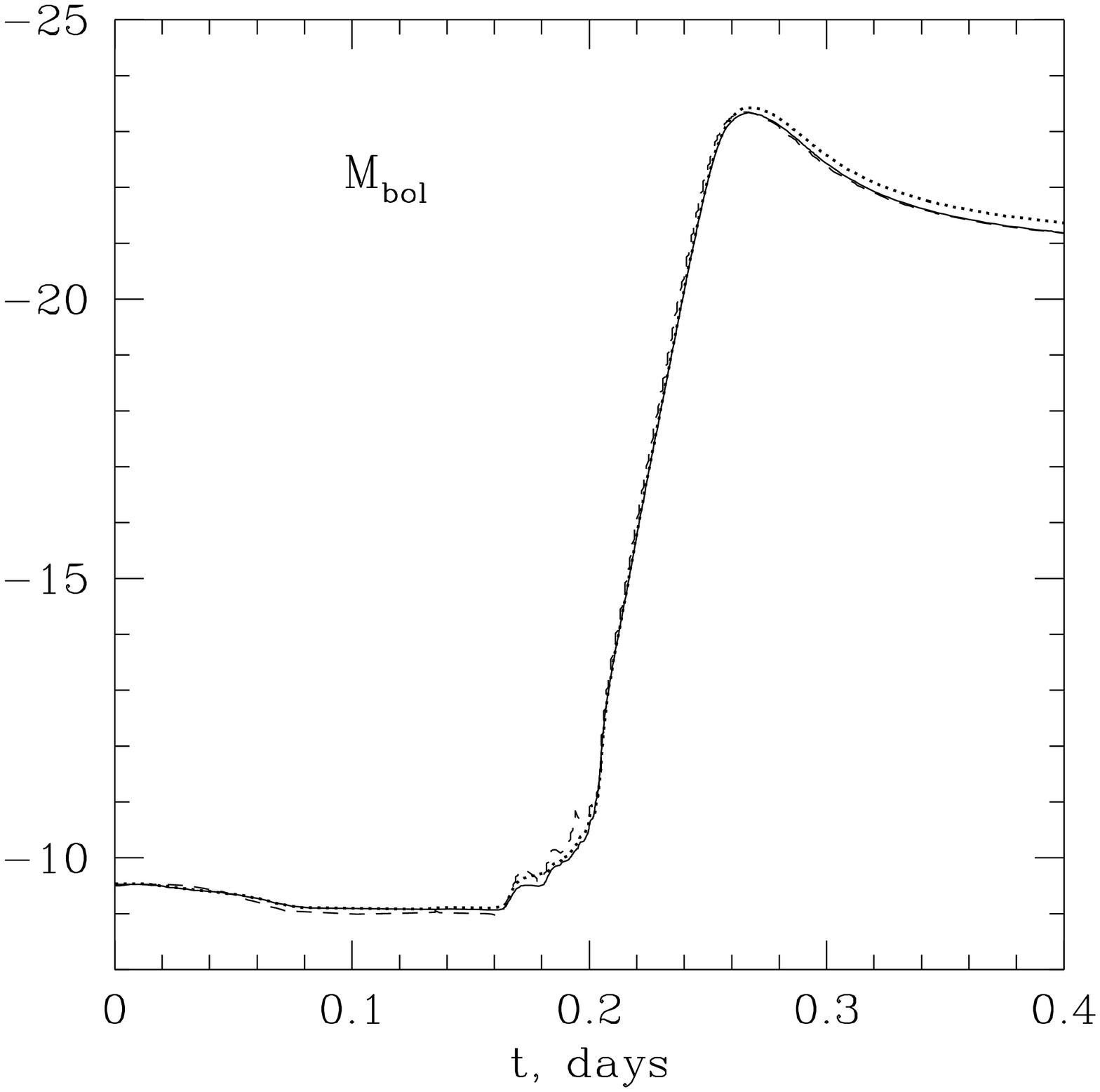}
\caption{Early bolometric light curve in the frame comoving with
         the outer radius of supernova.
         Dashed for Model 13C1 (scattering),
         solid for Model 13C2 (absorption).  Dots for the latter model
         in the obsever frame }
\label{earlum}
\end{figure}

\begin{figure}
\plotone{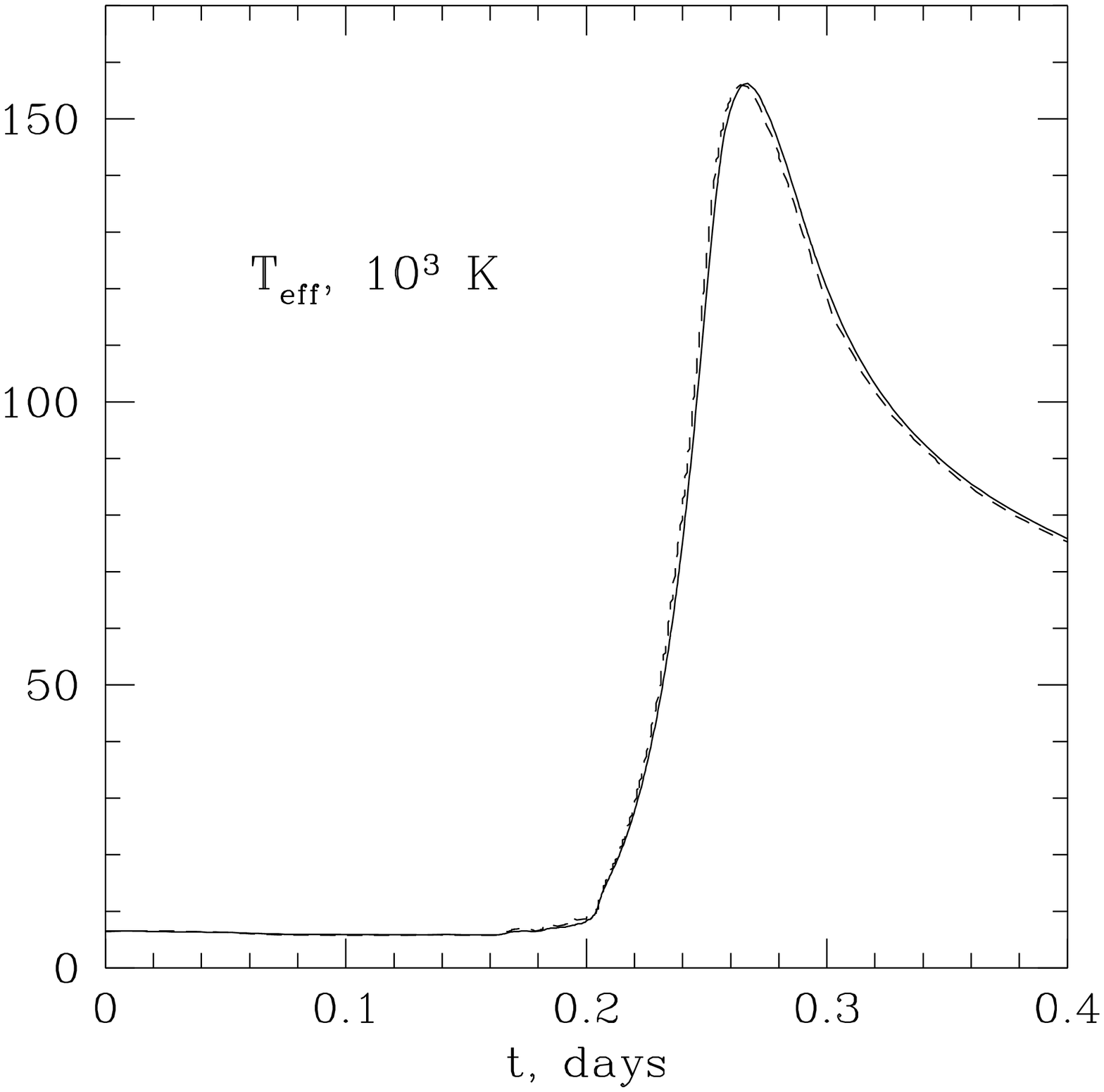}
\caption{Early $T_{\rm eff}$.
         Dashed for Model 13C1 (scattering),
         solid for Model 13C2 (absorption).
}
\label{earlte}
\end{figure}

Comparing the parameters near the first maximum found by KEPLER (Model
13B, Fig.16 of Woosley et al., 1994) and STELLA, we find that KEPLER
calculates a maximum luminosity $L=4\times10^{44}$ erg/s,
corresponding to $M_{\rm bol}=-22.8$ while STELLA finds $M_{\rm
bol}=-23.34$ (13C2) and -23.50 (13C1) (in the comoving frame). KEPLER
gives a maximum $T_{\rm eff}$ at 22000 sec of $\sim 130\times
10^3$ K and STELLA, $156.3\times 10^3$ K and $155.3\times 10^3$ K
respectively both at 0.267 d = 23070 sec.  Curves of $M_{\rm bol}(t)$
are given in Figure\ref{earlum} and of $T_{\rm eff}$ in
Figure\ref{earlte}.


It is not trivial to define an effective temperature for a
multi-frequency result.  STELLA uses the following algorithm for
finding $T_{\rm eff}$. At each time step and for any frequency we know
the flux $H_\nu$ and the Eddington factor, $h_\nu = H_\nu/J_\nu$, for
the outer boundary.  If we assume, that the intensity $I_\nu$ is
constant within a certain solid angle and zero outside (a hot spot
of uniform brightness), then $I_\nu=J_\nu/(1-h_\nu)$. For a blackbody
with $T_{\rm eff}$,
\begin{equation}
  aT_{\rm eff}^4 = \sum_\nu B_\nu dw_\nu
\label{sumbb}
\end{equation}
If we replace $B_\nu$ by the actual $I_\nu$ we find
\begin{equation}
  aT_{\rm eff}^4 = \sum_\nu J_\nu dw_\nu/(1-h_\nu)
\label{sumjj}
\end{equation}

With this definition we first define ``the last scattering
radius'', and then the $T_{\rm eff}$ from $L=\sigma
T_{\rm eff}^4 R_{\rm ph}^2$ relation. This $T_{\rm eff}$ it is not
necessarily associated with the temperature of any physical layer.  As
one can see in Figure \ref{earlte} this $T_{\rm eff}$ is practically
equal in models 13C1 and 13C2, that is for scattering and absorption
dominated opacity respectively. In the case of absorption this $T_{\rm
eff}$ is equal (to accuracy of order of one percent) to the matter
temperature at optical depth 2/3, but in the case of scattering the
matter at this depth is hotter by 20 -- 30 percent.

\begin{figure}
\plotone{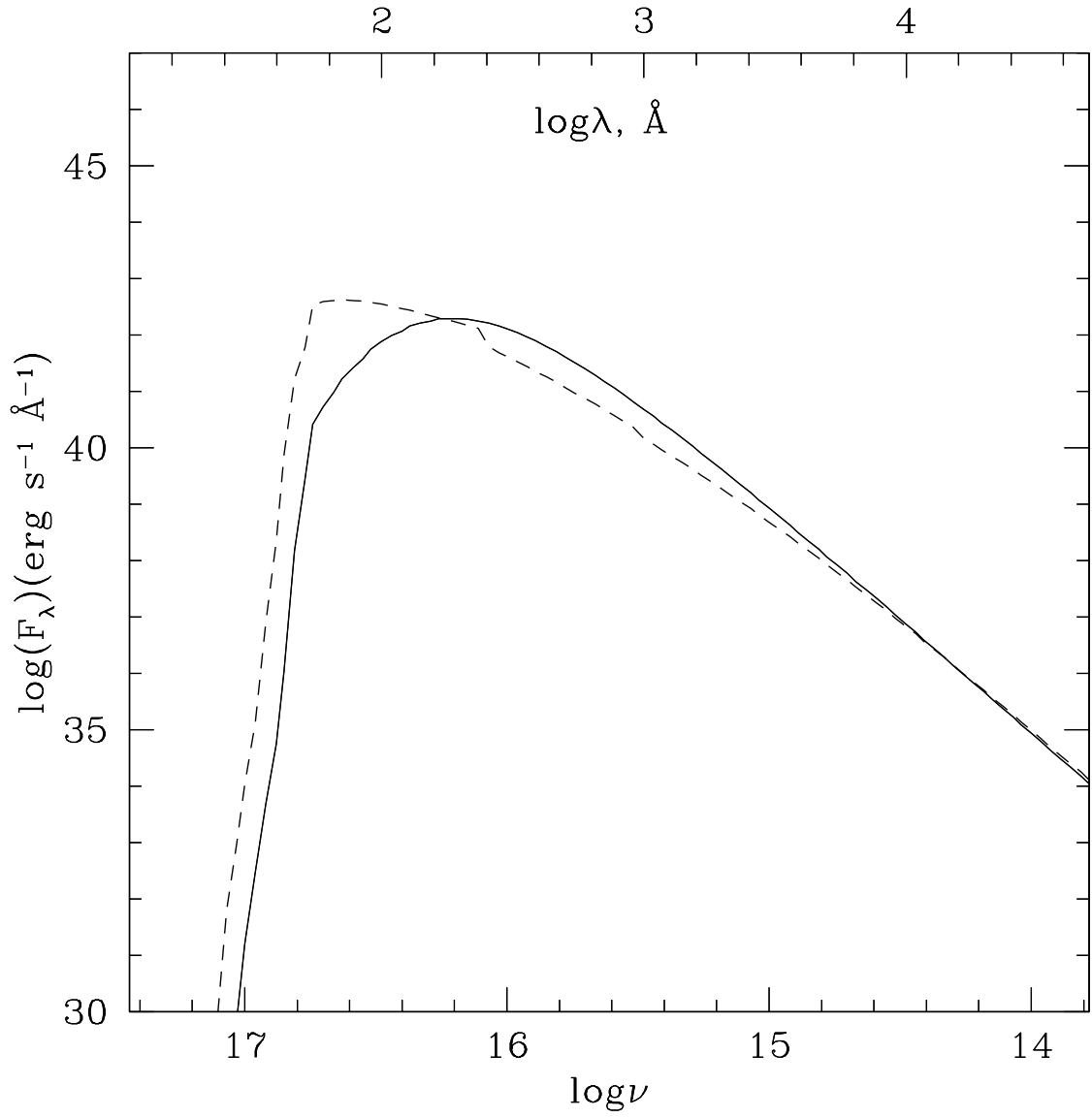}
\caption{Comoving luminosity at first maximum light for t=0.27 d.
         Dashed for Model 13C1 (scattering),
         solid for Model 13C2 (absorption).
}
\label{max1st}
\end{figure}

Figure \ref{max1st} gives the spectral luminosity for Models 13C1 and
13C2 around the first maximum at t=0.27 day. In the case of absorption
the spectrum is nearly a blackbody with a maximum $\lambda \simeq
200 \AA$. This gives a temperature $T_{\rm Wien} \simeq 1.5 \times
10^5$ K from the Wien displacement law. So in this case our $T_{\rm
eff}$ coincides with color and brightness temperatures over a large
part of the spectrum.  A different result holds for the scattering
case, Model 13C1. Now the spectrum deviates strongly from a blackbody
and the maximum at $\lambda \simeq 60 \AA$ gives $T_{\rm Wien} \simeq
4.8 \times 10^5 $, three times $T_{\rm eff}$. In the range of spectrum
corresponding to visual light both brightness and color temperatures
are {\em lower} than $T_{\rm eff}$ in the scattering model. For
example, $U-B=-1.2$ for 13C1 and $U-B=-1.3$ for 13C2 near the first
maximum light. Thus in visual light the scattering model might be
classified by observations as the cooler one which is opposite to
reality. This is simply the effect of dilution of a very hot radiation
born far below optical depth 2/3.

When discussing the second maximum of the light curve, we found that
Model 13C2 (fully absorptive opacity) was in better agreement both
with observations and with NLTE theory. At the second maximum the
actual opacity is dominated by lines and photoionization so Model 13C3
(absorptive lines, but realistic electron scattering) is practically
the same as 13C2 at that epoch. Here at the first maximum light the
situation is different. The opacity is overwhelmingly dominated by
electron scattering and now Model 13C1 is more physical.  For the
earliest epoch 13C3 is almost the same as 13C1, so the results
at shock breakout are not sensitive to the treatment of lines as
absorptive or scattering. 

Observers often estimate the photospheric temperature from blackbody 
spectral fits,
and this {\em color} temperature can be either higher or lower than
$T_{\rm eff}$ (Eastman, Schmidt, \& Kirschner 1996). If the blackbody
fit is good near the wavelength of peak emission, then the color
temperature is closer to the temperature at the
depth of thermalization, but in this case $R_{\rm ph}$ is neither the
surface of last scattering, nor the thermalization radius (for hot,
scattering dominated envelopes, it is much too small), so one must
exercise caution when comparing the theoretical predictions with the
`photospheric' parameters found by observers in these situations
(see Blinnikov \& Kozyreva 1998 for the analysis of the fits to 
early spectra of SN 1993J).


At late time $T_{\rm eff}$, as defined here, is much too low in
comparison to values given by observers, e.g. for 
$t=100$ days in the model 13C8 this definition gives 
$T_{\rm eff} \approx 2000$ K, while Richmond et al. (1994) give
blackbody $T \approx 5900$ K.
$T_{\rm eff}$ here is neither the color temperature, nor the matter
temperature at 
$\tau=2/3$ (for late times the total optical depth is too low and the
definition of equation  [\ref{sumjj}] looses its basic foundation: there is
no ``last scattering radius'', since only a small fraction of photons
experience a scattering).
We note, however, that $T_{\rm eff}$ 
is not a particularly convenient parameter for
comparison with observations.

\begin{figure}
\plotone{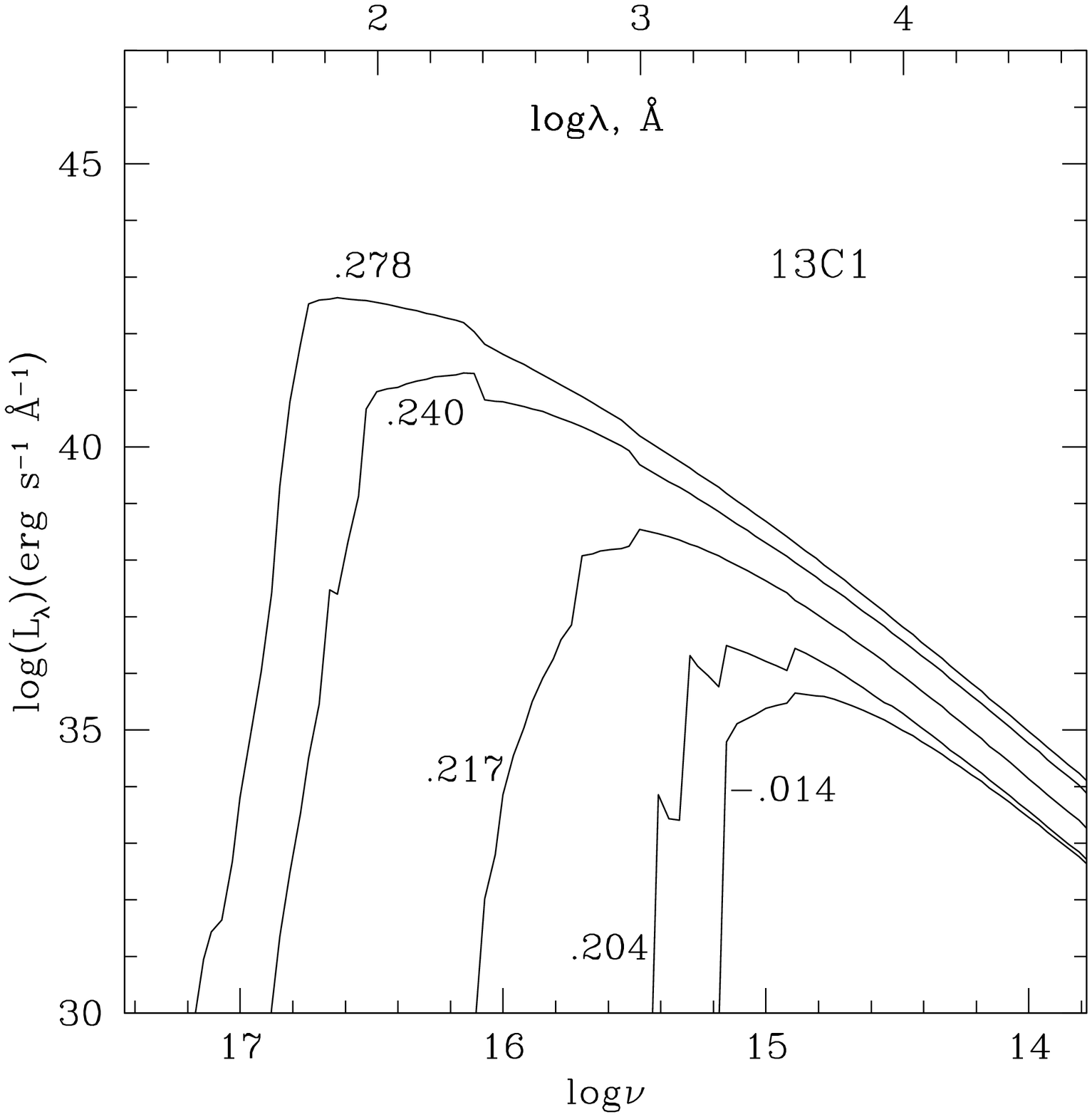}
\caption{Emerging spectral luminosity at the
 shock breakout in the observer frame for the run 13C1. 
 The curves are labeled by the retarded time
 in days.}
\label{flbreak1}
\end{figure}

\begin{figure}
\plotone{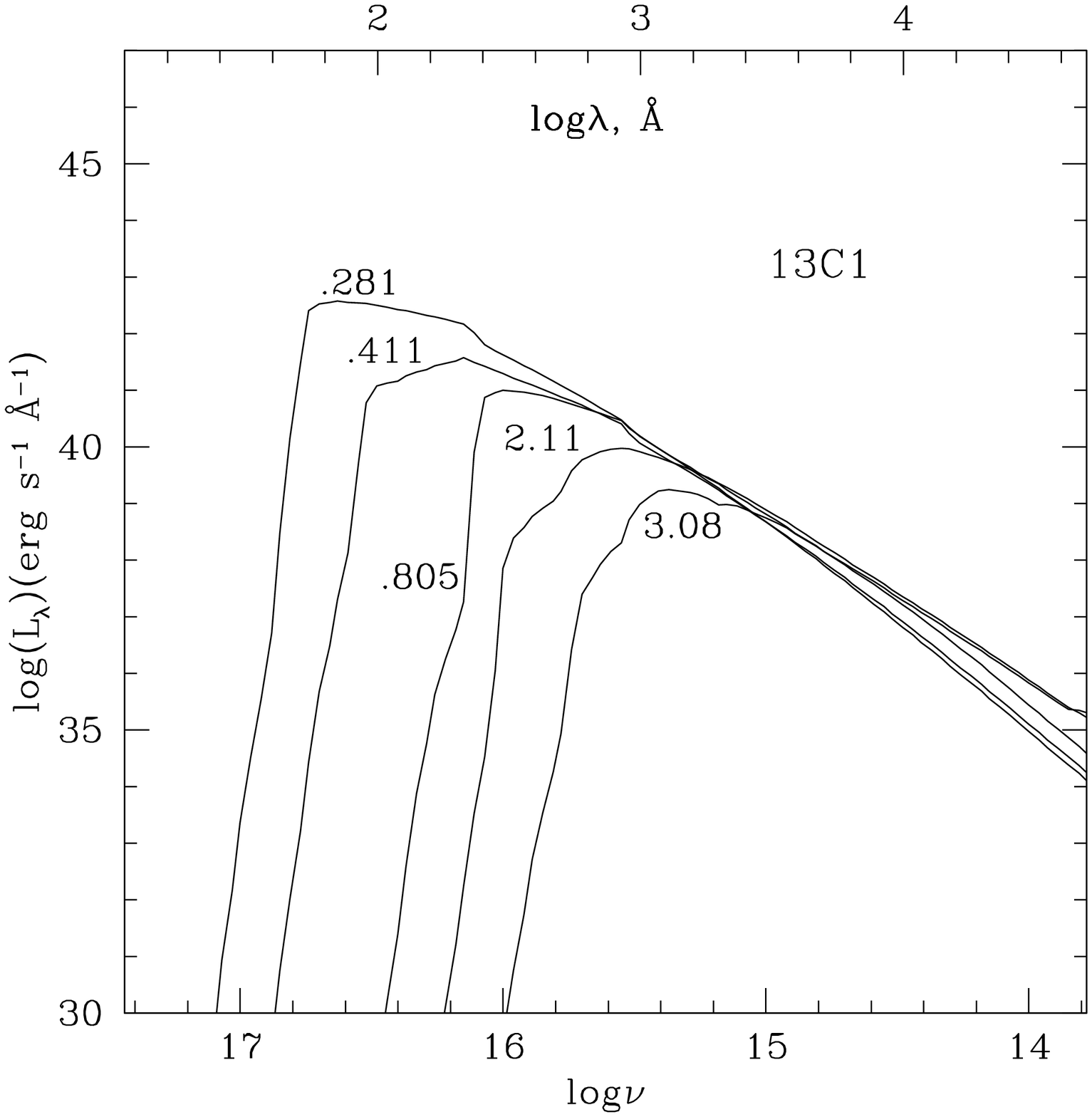}
\caption{The same as in Figure \protect\ref{flbreak1} but after the
shock breakout. }
\label{flbreak2}
\end{figure}

Figs \ref{flbreak1} and \ref{flbreak2}
display the outgoing spectrum for selected moments of time near the first
maximum.

Light curves for the case 
of pure scattering in lines (Model 13C1) and
the case of pure absorption in lines (Model 13C3) agree very well at
early times. This is because the opacity at these epochs is dominated
by electron scattering.

\section{CONCLUSIONS}

We have calculated light curves for Model 13C of Woosley et al (1994)
for SN 1993J using two treatments of radiation transport, EDDINGTON
(Eastman and Pinto 1993) and STELLA.  The explosion was calculated
with both STELLA and KEPLER. Good agreement was found between the two codes
for asymptotic profiles of density and velocity. We also found good
agreement between STELLA and EDDINGTON (and observations) for the
light curves and photometry, $M_{\rm bol}$ and $UBV$.

For times later than a few days after the second peak, the 
behavior of SN~1993J's
$U$ band light curve is in marked disagreement with either of the
two predicted $U$ light curves. A large disagreement is possible here
because only a small fraction of the light comes out in the ultraviolet.
The ultraviolet flux comes from a thin surface layer where the heating
and cooling are dominated by non-LTE processes not included in the
present calculation.

We have found that near the epoch of the second maximum of our light
curves additional cooling is necessary for better agreement with
observations and that the effects of expansion line opacity are very
important for the $U$ band.  EDDINGTON results and the $UBV$ fluxes of
STELLA sometime deviate substantially, especially at very late times,
i.e., on the tail of the light curve, but all those differences are within
reasonable limits given the uncertainty in opacity data and
non-equilibrium physics of late stages of light curves.

At shock breakout the uncertainties in opacity are not so important,
so one may hope to obtain even better predictions for an ultraviolet
flash at the shock breakout with a code like STELLA (where all
important effects in radiative transfer like aberration, Doppler shift
and time retardation are taken into account and are coupled to
hydrodynamics). At these stages the uncertainty of predictions must lie
in the uncertainty of the structure of the outermost layers of a
presupernova, in the presence of a superwind, or in effects of
asphericity.

Having calibrated the radiation hydro package STELLA both against
observations of SN 1993J and a more detailed numerical treatment of
just the radiation transport, we are now prepared to embark on
modeling the light curves of other Type II and eventually, Type I
supernovae. New predictions for the ultraviolet luminosity of SN 1993J
during its earliest moments are given in Figures \ref{flbreak1} -- 
\ref{flbreak2}

\begin{acknowledgments}

This work was begun in 
1993 during a visit by SB and OB to the UCSC campus. Its completion
was only made possible by help from many quarters.  We are very
grateful to Rob Hoffman and Marc Herant for their kind assistance and
very useful scientific discussions, to Dima Verner for providing his
atomic data in electronic form, and to the referee, Eddie Baron for
his valuable comments.  SB and OB thank David Koo for allowing them to
use his DEC-Alpha workstation and Robert Kurucz for providing his list
of lines on CDROM.  Part of this work was done during a visit by SB
and SEW stay at the MPA, Garching, and they thank Wolfgang Hillebrandt
for his hospitality.  Some calculations and a part of preparation of
the manuscript were done during SB visit to NAO, Mitaka, Tokyo, and to
Tokyo University, and he is grateful to Taka Kajino and to Ken Nomoto
for their warm hospitality and support and to Koichi Iwamoto and Tim
Young for helpful discussions.  The work was supported by the grants
from the California Space Institute (CS-58-93), the National Science
Foundation (AST-91-15367; AST 94-17161), NASA (NAG5 2843), and at
Lawrence Livermore National Laboratory by the US Department of Energy
(W-7405-ENG-48).  The work of SB and OB in Russia is supported by the
grants from the Russian Foundation for Fundamental Research
(96-02-16352, 96-02-17604) and from International Science \&
Technology Center 97-370.

\end{acknowledgments}

\appendix 

\section{Artificial radiative diffusion}

The system (\ref{comov}) -- (\ref{impul}) is 
of a hyperbolic type. The difference scheme used for its numerical 
solution with fluxes on
the mesh cell boundaries and energies in the cell centers has second 
order accuracy on a uniform grid and a low 
diffusivity, but does not take into account the hyperbolic nature
of equations. For hyperbolic systems other
schemes, those which make use of characteristic-ray solutions
of equations, are more  natural.
But it is their common property that when they are very stable, 
they are highly diffusive. As an example let us consider the simplest
equation 
\begin{equation}
   y_t + cy_x=0 \;,
\label{shype}
\end{equation}
where subscripts denote respective partial derivatives. The 
characteristic-ray of (\ref{shype}) is $x-ct={\rm const}$, so the solution
is $y=f(x-ct)$. Let $x_0$,
$x_1=x_0+\Delta x$, and $x_2=x_0+2\Delta x$ 
be the nodes of a uniform $x$-mesh. Then
the simplest first-order upwind approximation for $y_x$ at the point
$x_1$ is the  donor-cell one, 
\begin{equation}
   y_x=(y_1-y_0)/\Delta x \;,
\label{doncell}
\end{equation}
and this approximation gives good stable results in many practical
situation. Yet the more accurate, second order approximation to $y_x$ at
$x_1$,
\begin{equation}
   y_x=(y_2-y_0)/(2\Delta x )\;,
\label{secord}
\end{equation}
shows that in fact the donor-cell prescription (\ref{doncell}) approximates
to the second order in $\Delta x$ not the equation  (\ref{shype}), but
\begin{equation}
   y_t + cy_x=c\Delta x y_{xx} \;,
\label{difhype}
\end{equation}
that is a strong diffusion with the diffusivity $c\Delta x$. The
progress of numerical schemes in gas dynamics, e.g. 
the development of Godunov-type
approach, has successfully overcome the excessive diffusivity, but some
scheme diffusion is always needed for the stability of calculations.

We argue that in our case the diffusion is not only needed for
stability - it has a direct relevance to the description of the
correct propagation of radiative flux and energy for flashing surfaces.
Let us consider a simple problem (cf. Imshennik, Nadyozhin \& Utrobin 1981
where a more difficult problem is solved). 
Let a horizontal plane surface $S$ be dark
for time $t<0$ and bright with intensity $I_\nu$ constant in time and uniform 
in all directions for $t\ge 0$. If an observer
is above the plane $S$ at distance $z$ and there is pure
vacuum between the plane and the observer, then at the moment $t=z/c$
he sees not the whole bright surface $S$ but only a bright point directly
under him. For $t>z/c$ the point becomes a growing bright spot, and the
cosine of the angle to the edge of the spot relative the vertical 
direction is just $\mu_{\rm s}=z/ct$. Now we have for the angular moments
of intensity:

\begin{equation}
    J_\nu={1\over 2}\int_{z/ct}^1 I_\nu d\mu = 
      {I_\nu\over 2}\biggl(1-{z\over ct}\biggr)
\label{spotj}
\end{equation}

\begin{equation}
    H_\nu={1\over 2}\int_{z/ct}^1 I_\nu\mu d\mu = 
        {I_\nu\over 4}\biggl(1-{z^2\over c^2t^2}\biggr)
\label{spoth}
\end{equation}

\begin{equation}
    K_\nu={1\over 2}\int_{z/ct}^1 I_\nu\mu^2 d\mu = 
        {I_\nu\over 6}\biggl(1-{z^3\over c^3t^3}\biggr)
\label{spotk}
\end{equation}

We see that the flux and energy of radiation grow very smoothly,
they are smeared by the retardation, whereas the Eddington factor
$f_{\rm E}$ jumps from initial $1/3$ (for initial low but non-zero
brightness) to $1$ at the moment $t=z/c$ and later it relaxes back
to $1/3$ according to the relation
\begin{equation}
    f_{\rm E}={1\over 3}\biggl(1+{z\over ct}+{z^2\over c^2t^2}\biggr)
\label{eddrelx}
\end{equation}
To calculate the $f_{\rm E}$ jump very accurately is not an easy task, but
our goal is not $f_{\rm E}$ but the flux and the energy and they behave
quite smoothly.

That is, contrary to gas dynamics, with its shock waves, instantaneous
flashes on a star surface do not produce (in exact solutions) any
jumps in $J$ or $H$, due to time retardation. Thus an additional
diffusivity in this case does not destroy the correct solution.
The explicit form of the stabilizer is presented by the following expression
\begin{equation}
\dot{\cal H}_{\nu_{\rm diff}}=
      cR_{\rm vis}{{\cal H_\nu}_0 r^2_0 -2{\cal H_\nu}_1 r^2_1
                   +{\cal H_\nu}_2 r^2_2 \over 2 \Delta r\, r^2_1} \; .
\label{hstabl}
\end{equation}

The subscripts $0$, $1$, $2$ denote here the numbers of subsequent
radial zones.
The coefficient $R_{\rm vis}$, the `artificial radiative viscosity',
is of order unity in the transparent zones (then
the diffusivity of the donor cell scheme (\ref{difhype}) is reproduced there) 
and should be put equal to zero
for large optical depth. As our experiments show, the exact way of doing this
is not decisive, since the stabilizer (\ref{hstabl}) 
is important only at transient 
stages of fast variations of flux and is zero for cases when the luminosity
does not depend on radius. Moreover, there is no artificial viscosity added
to the zeroth moment (monochromatic radiative energy) 
equation (\ref{comov}) which actually governs the time behavior of the 
solution.

\section{Acceleration of mixing} 
 
There is some ambiguity in the treatment of large density contrasts
developing in the one-dimensional modeling of shock propagation
in radiating fluids (cf.  Falk \& Arnett 1977). After experimenting
with various prescriptions, the following expression for
$a_{\rm mix}$ at point 1 (for mesh points numbered 0, 1, 2)
was found to produce satisfactory results:
\begin{eqnarray}
 a_{\rm mix} = R_{\rm cut}(\tau_2)
   u_2\min(0,\hbox{div}\, u |_{3/2})-\nonumber\\
    - R_{\rm cut}(\tau_1) u_0\min(0,\hbox{div}\, u |_{1/2})  \;.
\label{mixac}
\end{eqnarray}
Here the cutting factor $R_{\rm cut}$ provides a normalization
of the effect of the artificial acceleration, is equal to zero
in the outermost zone, and  kills $a_{\rm mix}$ at optical
depths $\tau > 1$. The expression (\ref{mixac}) looks very much like
a gradient of a viscous pressure, but  it is constructed in such a way, 
that in effect
it does not change the kinetic energy integral. One can easily verify this 
using the relation $2u_i a_i=2u_i\dot u_i=d u_i^2/dt$ and the
expression (\ref{mixac}) for $a_i$ and summing over all zone numbers $i$.
Thus $a_{\rm mix}$ only redistributes the kinetic energy between
neighboring mass shells, where there is a strong compression, i.e.
$\hbox{div} u $ is negative, and, contrary to the artificial viscosity
$q$, one may not include any
effect associated with $a_{\rm mix}$ into the energy equation.

\end{document}